\documentclass[twocolumn]{aastex63}
\usepackage{graphicx}
\usepackage{amsmath}
\usepackage{cases}
\usepackage{color}

\DeclareMathAlphabet{\mathsc}{OT1}{cmr}{m}{sc}
\def\testbx{bx}%
\DeclareRobustCommand{\ion}[2]{%
\relax\ifmmode
\ifx\testbx\f@series
{\mathbf{#1\,\mathsc{#2}}}\else
{\mathrm{#1\,\mathsc{#2}}}\fi
\else\textup{#1\,{\mdseries\textsc{#2}}}%
\fi}

\newcommand{\beq}{\begin{equation}}
\newcommand{\eeq}{\end{equation}}

\newcommand{\hi}{\ion{H}{i}~}
\newcommand{\HI}{\ion{H}{i}}
\newcommand{\kms}{km ${\rm s^{-1}}$~}
\newcommand{\kmsa}{km ${\rm s^{-1}}$}

\newcommand{\nhi}{$\mathrm{N}_\mathrm{H{\sc I}}$}
\newcommand{\nHI}{$\mathrm{N}_\mathrm{H{\sc I}}~$}

\shortauthors{Clark \& Hensley}

\begin{document}

\title{Mapping the Magnetic Interstellar Medium in Three Dimensions\\ Over the Full Sky with Neutral Hydrogen}

\author[0000-0002-7633-3376]{S.E. Clark}
\affiliation{Institute for Advanced Study, 1 Einstein Drive, Princeton, NJ 08540, USA}
\altaffiliation{Hubble Fellow}
\email{seclark@ias.edu}

\author[0000-0001-7449-4638]{Brandon S. Hensley}
\affiliation{Spitzer Fellow, Department of Astrophysical Sciences, Princeton University, Princeton, NJ 08544, USA}

\begin{abstract}
Recent analyses of 21-cm neutral hydrogen (\HI) emission have demonstrated that \hi gas is organized into linear filamentary structures that are preferentially aligned with the local magnetic field, and that the coherence of these structures in velocity space traces line-of-sight magnetic field tangling. On this basis, we introduce a paradigm for modeling the properties of magnetized, dusty regions of the interstellar medium, using the orientation of \hi structure at different velocities to map ``magnetically coherent'' regions of space. We construct three-dimensional (position-position-velocity) Stokes parameter maps using \HI4PI full-sky spectroscopic \hi data. We compare these maps, integrated over the velocity dimension, to \textit{Planck} maps of the polarized dust emission at 353\,GHz. Without any free parameters governing the relation between \hi intensity and dust emission, we find that our $Q$ and $U$ maps are highly correlated ($r > 0.75$) with the 353\,GHz $Q$ and $U$ maps of polarized dust emission observed by {\it Planck} and reproduce many of its large-scale features. The $E/B$ ratio of the dust emission maps agrees well with the \HI-derived maps at large angular scales ($\ell \lesssim 120$), supporting the interpretation that this asymmetry arises from the coupling of linear density structures to the Galactic magnetic field. We demonstrate that our 3D Stokes parameter maps constrain the 3D structure of the Galactic interstellar medium and the orientation of the interstellar magnetic field.  
\end{abstract}


\section{Introduction}

The Galactic magnetic field pervades the interstellar medium (ISM) and affects myriad physical processes, including cosmic ray propagation, gas dynamics, and the formation of stars and molecular clouds. Despite its importance, the structure of the Galactic magnetic field is poorly understood, in part because of inherent limitations in observational tracers: the data tend to probe only a single component of the 3D magnetic field, a single phase of the multi-phase ISM, and/or line-of-sight integrated averages of magnetic properties \citep[e.g.,][]{Ferriere:2001}. Polarized light from the magnetic ISM is also a complex foreground for observations of the polarized Cosmic Microwave Background (CMB), thereby tethering cosmological pursuits to our understanding of interstellar magnetism. A complete picture of the three-dimensional structure of the interstellar magnetic field is both a formidable challenge and a worthwhile pursuit \citep{Haverkorn:2015}.  

Polarized emission in the far infrared (FIR) is dominated by thermal emission from rotating dust grains that align with their short axes preferentially parallel to the local magnetic field. Measurements of the polarized FIR emission thus trace the magnetic field orientation in the dust, projected onto the plane of sky and integrated along the line of sight. The Galactic polarized dust emission was recently mapped at 353 GHz over the full sky by the \textit{Planck} satellite \citep{PlanckXIX}, enabling better characterization of the polarized CMB foreground, as well as studies of the magnetic, turbulent ISM \citep{Planck2018XI, Planck2018XII}. 

A principal limitation of dust emission as a probe of the magnetic field is that it is necessarily an integrated measure. The three-dimensional structure of the ISM, and variations in the magnetic field orientation along the line of sight, cannot be directly measured from FIR emission. The line-of-sight magnetic structure of the dusty ISM is thus poorly constrained from the data, despite its importance both for ISM studies and for cosmological foregrounds. 

Unlike measurements of the FIR dust continuum, spectroscopic observations of 21-cm neutral hydrogen (\HI) line emission provide information in three dimensions: position-position-velocity, where the third dimension is the line-of-sight velocity associated with the shift from the line rest frequency ($v_{lsr}$). High-resolution channel maps of this \hi emission reveal slender linear features, well aligned with the local magnetic field orientation \citep{Clark:2014, Clark:2015}. The dispersion in the orientation of these structures along the velocity dimension is correlated with the fractional polarization of dust emission, suggesting that \hi can probe not only the plane-of-sky magnetic field orientation, but also the magnetic ``tangling" along the line of sight \citep{Clark:2018}. We combine these insights, along with the fact that gas and dust are generally well-mixed in the diffuse ISM \citep[e.g.,][]{Lenz:2017}, to define the magnetic coherence of regions of the neutral ISM.

This paper is organized as follows. In Section~\ref{sec:data} we introduce the data used in this work. In Section~\ref{sec:model} we set forth our principle of magnetic coherence, its observational motivation, and its application to the maps derived in this work. In Section~\ref{sec:3dmaps} we present three-dimensional Stokes parameter maps over the full sky. In Section~\ref{sec:galfacomp} we compare those maps with partial-sky maps derived from higher-resolution \hi data. In Section~\ref{sec:planckcomp} we compare our full-sky maps to \textit{Planck} 353\,GHz polarized dust observations, including derived properties like the polarization fraction, polarization angle dispersion function, and $E$- and $B$-mode cross-power spectra. In Section~\ref{sec:otherdatacomp} we compare our three-dimensional Stokes maps to other tracers of the magnetized ISM in selected regions of sky: low-frequency radio polarimetric observations and starlight polarization measurements. We discuss variations and possible extensions to the methods presented here in Section~\ref{sec:variations}. In Section~\ref{sec:discussion} we discuss the utility of our maps for cosmology and ISM structure, and next steps in higher-dimensional mapping of the magnetic ISM. We summarize and conclude in Section~\ref{sec:conclusions}.

\section{Data}
\label{sec:data}
\subsection{Neutral Hydrogen}
Neutral hydrogen is observed via the $\lambda$21-cm spin-flip transition, and is a ubiquitous tracer of interstellar gas: no known sightline lacks Galactic \hi emission. Recent technological advances have enabled high dynamic range observations of the \hi sky. The Leiden/Argentine/Bonn Survey \citep[LAB;][]{Kalberla:2005}, long the gold standard full-sky \hi survey with angular resolution $\vartheta_{fwhm}=36'$, was recently superseded in both sensitivity and resolution by the \hi $4\pi$ Survey \citep[\HI4PI;][]{HI4PI:2016}. \HI4PI, with angular resolution $\vartheta_{fwhm}=16.2'$ and spectral resolution $\delta v = 1.49$ \kmsa, is the highest-resolution full-sky \hi survey to date, achieved by combining the Effelsberg-Bonn \hi Survey of the northern sky \citep[EBHIS;][]{Winkel:2016} with the Parkes Galactic All-Sky Survey in the south \citep[GASS;][]{McClure-Griffiths:2009}. The Galactic Arecibo L-Band Feed Array \hi Survey \citep[GALFA-\HI;][]{Peek:2018} is a higher-resolution ($\vartheta_{fwhm}=4.1'$, $\delta v = 0.184$ \kmsa) survey of the entire sky visible to the 305-m Arecibo telescope, $\sim32\%$ of the celestial sphere. For a comprehensive comparison of modern large-area \hi surveys, see \citet{Winkel:2016}. 

In this work we make use of the alignment between slender \hi features in narrow spectral channels and the plane-of-sky magnetic field \citep{Clark:2015}. Analysis of both GASS and GALFA-\hi data found that the alignment between \hi features and the projected magnetic field is sharper for higher spatial resolution \hi data \citep{Clark:2014}. For our purposes, this amounts to a trade-off between the superior angular resolution of GALFA-\hi and the optimal sky coverage of \HI4PI. This work focuses on \HI4PI data, in order to study the 3D distribution of the magnetic ISM over the full sky, but we also build maps from GALFA-\HI, and compare the two in Section~\ref{sec:galfacomp}.

\begin{figure}
\centering
\includegraphics[width=\columnwidth]{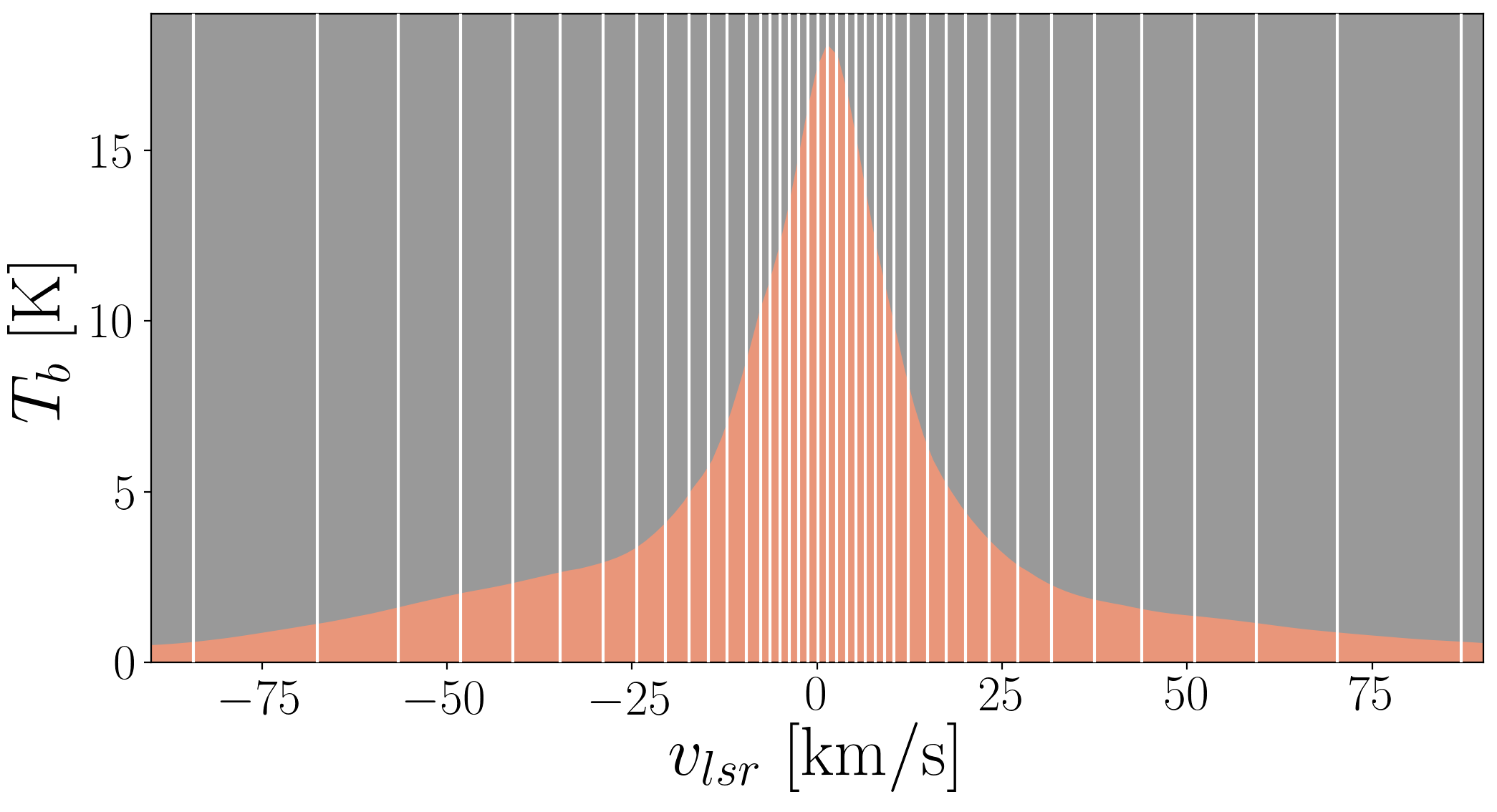}
\caption{\hi velocity channels used in the \HI4PI analysis. Velocity channel bins are plotted over the mean $T_b(v)$ spectrum over the full \HI4PI map. }\label{fig:HIbinning}
\end{figure}

We bin the \HI4PI data into velocity channels such that there is approximately equal integrated intensity in each pair of channels moving symmetrically outward from the center line. This binning scheme, illustrated in Figure~\ref{fig:HIbinning}, uses the native $\Delta v = 1.3$\,\kms channel width for velocity channels near $v_{lsr} = 0$, and channel widths of up to 20.6\,\kms in the outermost bins. This velocity channel binning is similar to the one used by \citet{Lenz:2019}. Throughout this work, we denote the integrated \ion{H}{i} intensity in a velocity bin centered on velocity $v$ with width $dv$ as $I\left(v\right)$, i.e.,

\begin{equation}
    \label{eq:IHI_def}
    I\left(v\right) = \int_{v-dv/2}^{v+dv/2} T_b\left(v'\right)dv'
    ~~~,
\end{equation}
where $I\left(v\right)$ is measured in K\,km\,s$^{-1}$. We also define the \hi column density

\beq\label{eq:NHI}
N_{\HI} = 1.82 \times 10^{18} \int_{-90}^{+90} T_b\left(v\right)dv
~~~,
\eeq
where $N_{\HI}$ is in $\mathrm{cm}^{-2}$ for $T_b$ in K and $v$ in \kmsa, and the integral over -90 \kms $< v <$ +90 \kms is chosen so that $N_{\HI}$ mostly represents Galactic emission. Equation~\ref{eq:NHI} is correct under the assumption that the \hi emission is optically thin, which is not valid for all sightlines in these data, but is generally a good approximation at high Galactic latitudes \citep{Murray:2018}.

We also use the publicly available \hi intensity and RHT output of $\Delta v = 3.7$\,\kms GALFA-\hi channel maps from -36.4\,\kms to +37.2\,\kms described in \citet{Peek:2018}, and convert to Galactic coordinates. Maps constructed from GALFA-\hi data not only use independent 21-cm observations that are higher resolution than \HI4PI, but also a different velocity range and binning scheme. The GALFA-\hi observations span $-1^\circ \lesssim$ decl. $\lesssim +38^\circ$, but we make a conservative cut on the area considered in this work to $1.5^\circ <$ decl. $< 35.5^\circ$ to avoid telescope scan artifacts at the edges of the Arecibo declination range.

\subsection{Dust Emission}
\label{subsec:data_dust}

We make use of several data products released by the \textit{Planck} collaboration for comparison with our \HI-based maps. We use the $80'$ R3.00 $353$\,GHz Stokes $I$, $Q$, and $U$ maps post-processed with the Generalized Needlet Internal Linear Combination algorithm \citep[GNILC;][]{Remazeilles:2011} to remove the anisotropic Cosmic Infrared Background (CIB) from the Galactic dust emission. Following the fiducial offset correction adopted by the \textit{Planck} collaboration, we subtract 452\,$\mu \mathrm{K}_{\mathrm{CMB}}$ from the GNILC Stokes $I$ map to correct for the CIB monopole at 353\,GHz. We also add a Galactic offset correction of $63$\,$\mu\mathrm{K}_{\mathrm{CMB}}$ to the $I$ map \citep[see discussion in][]{Planck2018XII}. We compute the modified asymptotic estimator introduced by \citet{Plaszczynski:2014} to create maps of the noise-debiased polarization fraction. The noise debiasing is a minor adjustment at $80'$ resolution. Hereafter we refer to this $80'$, debiased, offset-corrected, GNILC-derived polarization fraction map as $p_{353}$. 

For the power spectra comparisons in Section~\ref{subsec:cross_spectra} we employ the 353\,GHz R3.01 half-mission $I$, $Q$, and $U$ maps. We subtract the CMB contribution from these maps using the R3.00 \texttt{SMICA} CMB maps, and then smooth to $16.2'$ to match the native resolution of the \HI4PI data.

\section{Model}
\label{sec:model}

\subsection{Observational basis}\label{sec:obsbasis}

In this work we present a framework for mapping the three-dimensional structure of the magnetic neutral ISM using spectroscopic observations of the 21-cm line. We rely on three observational facts that relate the structure of \hi to the properties of dust emission. These are as follows:

\begin{enumerate}
\item The \hi column density traces dust column density in the diffuse ISM \citep[e.g.,][]{Lenz:2017}.
\item \hi features in narrow spectral channel maps are aligned with the plane-of-sky magnetic field \citep{Clark:2015}.
\item The coherence of \hi orientation as a function of velocity traces the degree of line-of-sight magnetic field tangling \citep{Clark:2018}.
\end{enumerate}
 
Dust and gas are well-mixed in the diffuse ISM as evidenced by the tight empirical correlation between the \hi column density (\nhi) and the dust extinction and emission \citep{Burstein:1978,Boulanger:1996}. Although this relationship breaks down at higher column densities where molecular gas can constitute a significant fraction of the hydrogen column, \nHI is a reliable tracer of the total dust column for low-column lines of sight. Recent analysis of the HI4PI data found that \nHI and the dust reddening $E(B-V)$ are well-fit by a linear relationship for \nHI $< 4\times10^{20}$\,cm$^{-2}$ with a scatter of $\sim 10\%$ \citep{Lenz:2017}.

High-resolution images of \hi emission display prominent linear structure that is well aligned with the plane-of-sky magnetic field, as traced by both optical starlight polarization \citep{Clark:2014} and FIR polarized dust emission \citep{Clark:2015, Martin:2015}. Magnetically aligned \hi intensity structures are a ubiquitous feature of the diffuse ISM. The aligned structures are thought to be components of the cold neutral medium (CNM) on the basis of temperature estimates from linewidth measurements \citep{Clark:2014, Kalberla:2016}, the existence of similar structures observed in \hi absorption in the Galactic plane \citep{McClure-Griffiths:2006}, an enhanced FIR/\nHI ratio at the locations of these structures in the diffuse ISM \citep{Clark:2019}, and the dependence of the equivalent width of interstellar \ion{Na}{i} absorption on the column density in small-scale channel map structures \citep{PeekClark:2019}. 

Similar alignment between diffuse FIR dust intensity structures and the projected magnetic field was measured by {\it Planck} \citep{PlanckXXXII}. This alignment is thought to give rise to some of the statistical properties of the \textit{Planck} polarized dust emission, notably the positive $TE$ correlation and the non-unity $EE/BB$ ratio \citep{PlanckXXXVIII}. Indeed, template $E$- and $B$-mode maps constructed from \hi orientation alone (with no information on the polarized intensity) measure $EE/BB \sim 2$ in the diffuse ISM, in rough agreement with the \textit{Planck} measurement \citep{Clark:2015}. If the $E/B$ asymmetry and positive $TE$ correlation are a consequence of the alignment of density structures and the magnetic field, we expect that our maps will naturally reproduce these statistical properties \citep[see also][]{Ghosh:2017,Adak:2019}. 

The final observational point allows us to model variable magnetic field orientations along the line of sight in a data-driven manner. \citet{Clark:2018} computed a metric, termed ``\hi coherence,'' quantifying the degree of order or disorder in \hi orientation in different velocity channels along the same line of sight. Lines of sight where the orientations of \hi features are relatively aligned across velocity channels correspond to an ordered field. Conversely, sightlines with relatively misaligned \hi features in different velocity channels correspond to sightlines probing multiple field orientations. Indeed, the \hi coherence was found to be a strong predictor of the \textit{Planck} 353\,GHz fractional polarization ($p_{353}$) in the diffuse ISM. 

We emphasize that \hi LOS velocity is not a one-to-one probe of distance, particularly at high Galactic latitude where the \hi is not strongly sheared by Galactic rotation. If two differently-oriented \hi features lie at two velocities along a single line of sight, it is typically not possible to say from the \hi data alone where the two features exist along the distance axis. However, since \hi orientation traces the plane-of-sky magnetic field orientation, these two \hi features imply that two regions with differently-oriented magnetic fields lie somewhere along the line of sight. For the purposes of modeling the magnetic field in the neutral medium traced by polarized dust emission, it is enough to partition the \hi intensity into regions with distinct magnetic field orientations.

\begin{figure*}
\centering
\includegraphics[width=0.9\textwidth]{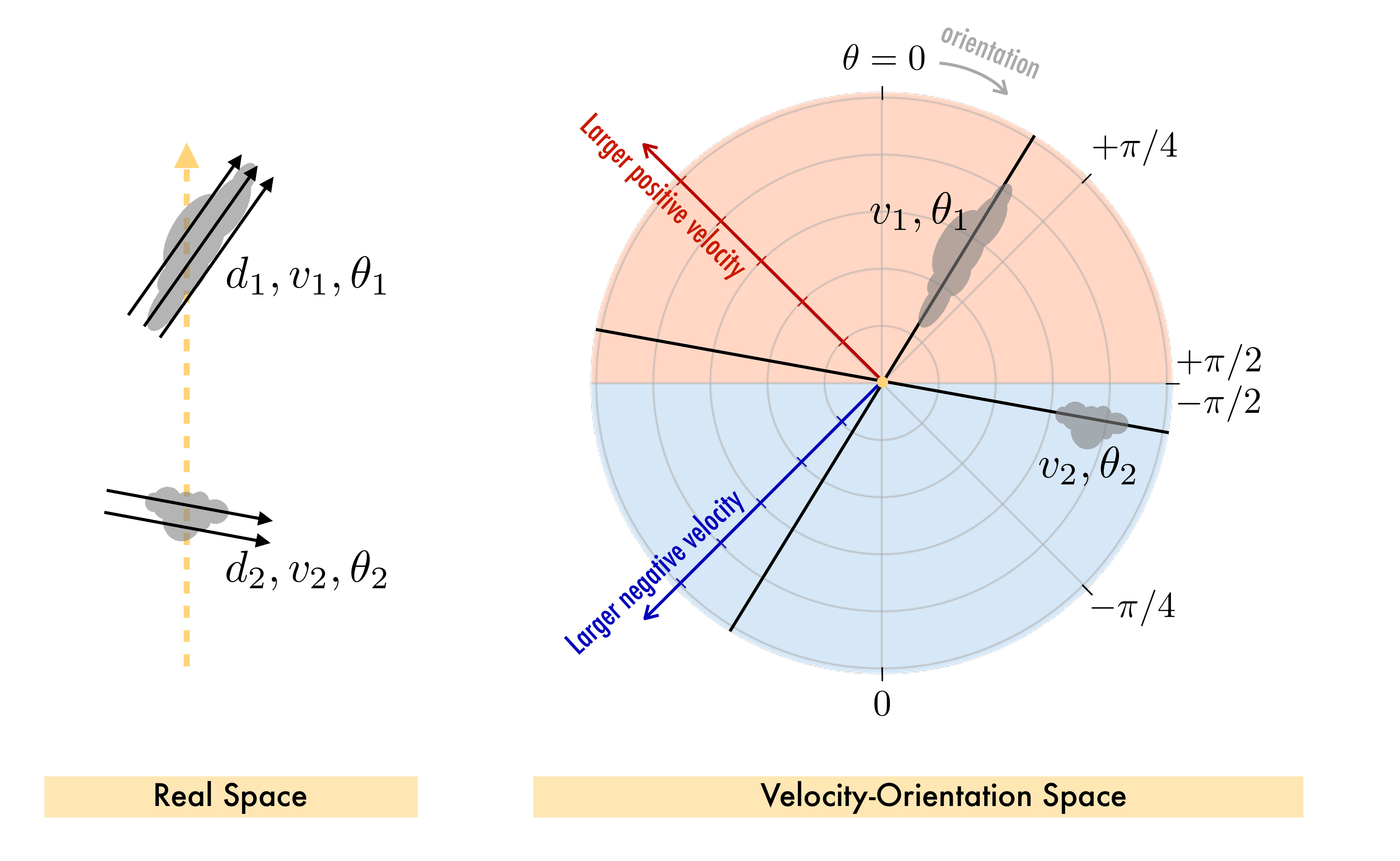}
\caption{Diagram of the magnetically coherent cloud paradigm. Left: ``real space" depiction of dust clouds along a line of sight. The clouds are at distances $d_1$ and $d_2$ from the observer, and sit in local magnetic fields with plane-of-sky orientations $\theta_1$ and $\theta_2$. The \hi associated with each cloud has a median velocity $v_1$ and $v_2$, respectively. Right: clouds mapped into velocity-orientation space. This circular diagram represents the distribution of \hi data as a function of line-of-sight velocity and plane-of-sky orientation. The magnitude of the velocity increases radially outward from the center of the diagram, with positive velocities (relative to $v_{lsr}$) plotted on the upper hemisphere (red), and negative velocities plotted on the lower hemisphere (blue). Plane-of-sky orientation is plotted azimuthally on [-$\pi/2$, $+\pi/2$), respecting the 180$^\circ$ degeneracy in orientation angle. Solid black lines denote the orientations $\theta_1$ and $\theta_2$. Velocity-orientation space separates the data into magnetically coherent regions: even in the case that $v_1 = v_2$, the two clouds would occupy different regions in this diagram.}\label{fig:diagram}
\end{figure*}

\begin{figure*}
\centering
\includegraphics[width=\textwidth]{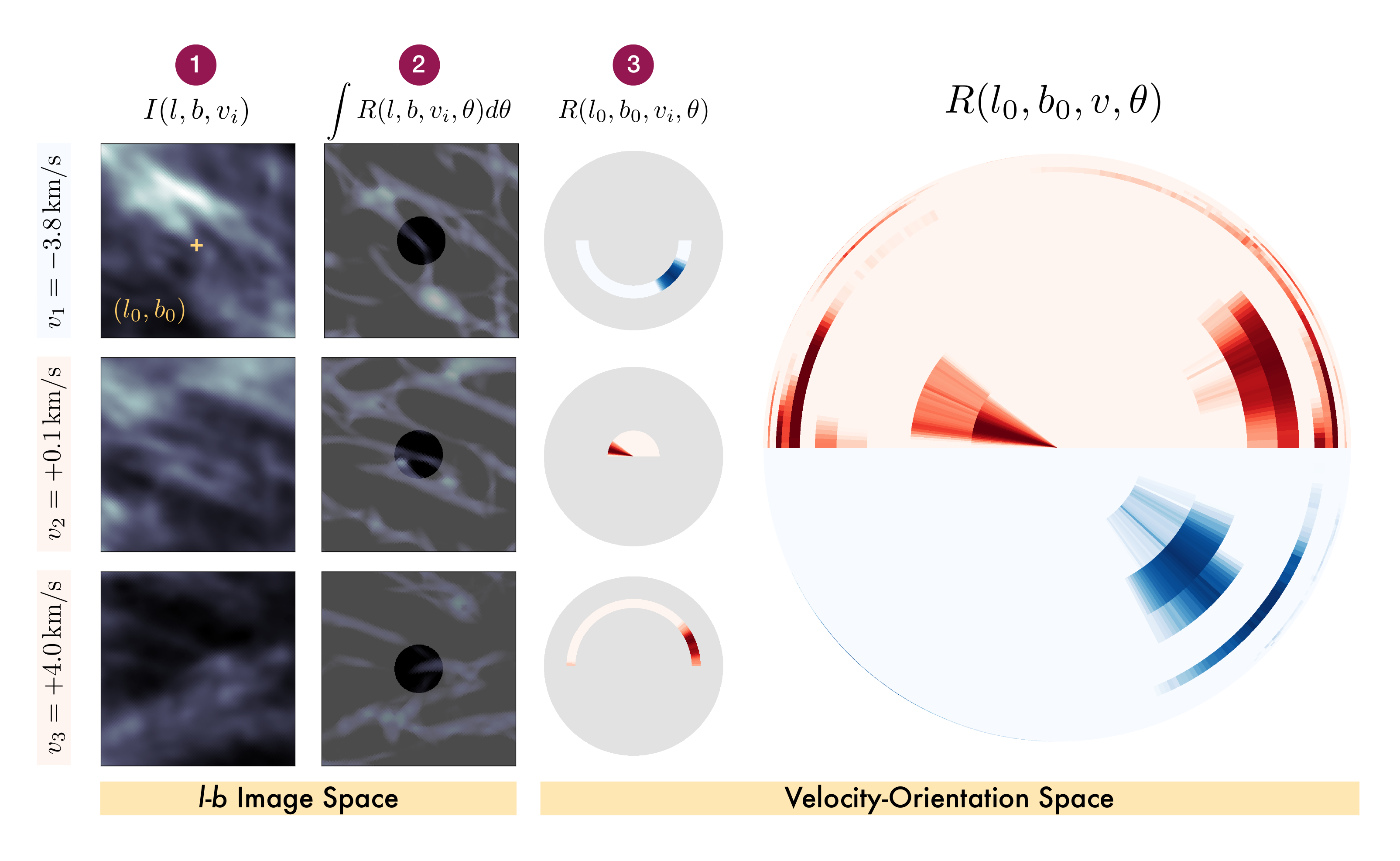}
\caption{The anatomy of a single sightline in velocity-orientation space. 1. Lefthand column shows the \hi intensity in a single velocity channel, $I(l, b, v_i)$ in a $4^\circ \times 4^\circ$ region of sky centered on $(l_0, b_0) = (115.5^\circ, -25.3^\circ)$. Rows show this region in three velocity bins: $v_1$, $v_2$, and $v_3$, centered on $-3.8$ \kmsa, $+1.0$ \kmsa, and $+4.0$ \kmsa, respectively. 2. Second column shows the total linear intensity, or RHT backprojection, in the same region of sky: $\int R(l, b, v_i, \theta)d\theta$. The circular transparent region in each of these thumbnails highlights the approximately $75'$ diameter region over which RHT orientation is calculated for the center pixel $(l_0, b_0)$. 3. Third column shows RHT output $R(l_0, b_0, v_i, \theta)$ -- that is, linear intensity as a function of orientation around $(l_0, b_0)$, plotted in $v-\theta$ space for a single velocity channel per row. These represent three velocity slices of the full $R(l_0, b_0, v, \theta)$ for this sightline plotted in the rightmost figure panel. This representation of the velocity-orientation space distribution of linear \hi features follows the coordinate system laid out in Figure~\ref{fig:diagram}. For visual clarity, each $(v, \theta)$ bin has roughly equal area. }\label{fig:ninepanel_RHT_v_theta}
\end{figure*}

\subsection{Magnetically coherent cloud paradigm}\label{sec:coherentclouds}

Based on the insights outlined in Section~\ref{sec:obsbasis}, we posit that \hi emission features that are coherent in velocity, plane-of-sky orientation, and spatial extent represent magnetically and spatially coherent clouds. This idea is shown schematically in Figure~\ref{fig:diagram}: two clouds along a single line of sight that sit in spatially distinct regions with differently oriented magnetic fields will map to different regions of velocity-orientation space. Observations of the 21-cm line emission (GALFA-\hi or \HI4PI) map the \hi intensity as a function of sky position and line-of-sight velocity, $I_\ion{H}{i}(l, b, v)$, where $(l, b)$ represent sky coordinates and $v$ represents the line-of-sight velocity. By measuring the orientation $\theta$ of linear structures in each velocity channel, we wish to map this emission into $(l, b, v, \theta$) space. The distribution of \hi emission in $(l, b, v, \theta)$ traces ``magnetic coherence'' without cloud-finding or otherwise prescribing dust properties in three-dimensional space.

Mapping $I_\ion{H}{i}(l, b, v)$ to $I_\ion{H}{i}(l, b, v, \theta)$ requires an algorithm that can measure the distribution of \hi as a function of orientation on the sky. We use the Rolling Hough Transform \citep[RHT;][]{Clark:2014} to quantify the coherent linearity of \hi emission in each channel map. The RHT maps image-plane data into position-position-orientation space by 1. high-pass filtering and bitmasking the image to highlight structure on small spatial scales, 2. selecting a circular window of data with diameter $D_W$ centered on each image pixel, 3. calculating the fraction of the window that is nonzero in the bitmask as a function of orientation, $R(\theta)$, and 4. thresholding $R(\theta)$ at a pre-specified fraction of the window length, $Z$. The algorithm used for step 3 is based on the Hough transform \citep{Hough:1962tb}. The basic principle is that straight lines in $(x, y)$ image space are represented by points in $(\rho, \theta)$ space, given the mapping
\beq
\rho = x\,\mathrm{cos}\theta + y\,\mathrm{sin}\theta
~~~.
\eeq
In the RHT, each $(x, y)$ image position within a given circular window is mapped to the space of possible lines with $\rho=0$: that is, lines that pass through the center of the window. 

The high-pass filtering step of the RHT is equivalent to an unsharp mask. In our analysis of the \HI4PI data we use a Gaussian filter with FWHM=$30'$, following \citet{Kalberla:2016}. We use $D_W = 75'$ and $Z=0.7$ for consistency with \citet{Clark:2015}. We use the canonical $\theta$-binning of \citet{Clark:2014}, but note that the alternative binning proposed by \citet{Schad:2017} does not substantially change the results. To avoid distortion effects from the map projection, we project each rolling window such that the center of the window is the center of the projection, as was done for the GALFA-\hi DR2 data in \citet{Peek:2018}. The RHT works on any two-dimensional data, and apart from \hi observations, it has been used to quantify the orientation of structures for such disparate applications as dust emission \citep{Malinen:2016}, the solar corona \citep{AsensioRamos:2017}, depolarization canals in radio polarimetric observations \citep{Jelic:2018}, X-ray data \citep{Marelli:2019}, and magnetohydrodynamic (MHD) simulations \citep{Inoue:2016wb}.

By performing the RHT on each velocity channel of the \hi data, we calculate $R(v, \theta)$, the linear intensity as a function of orientation and line-of-sight velocity. We normalize the RHT amplitude at each velocity $v$ such that
\beq
    \int R\left(v, \theta\right) d\theta = 1~~~.\label{eq:RHTnorm}
\eeq
We can therefore treat $R\left(v, \theta\right)$ analogously to a probability distribution function over the possible orientations $\theta$ in a given pixel at a given velocity. Note that Equation~\ref{eq:RHTnorm} only applies where the RHT output is nonzero: some velocity channels will have $\int R\left(v, \theta\right) d\theta = 0$ if no prominently linear structure is detected. 

The mapping to velocity-orientation space is the backbone of the magnetic coherence paradigm illustrated in Figure~\ref{fig:diagram}. We show an example of this mapping using the RHT applied to \HI4PI data in Figure~\ref{fig:ninepanel_RHT_v_theta}. 

With measurements of both $I\left(v\right)$ and $R\left(v, \theta\right)$, we can construct synthetic Stokes parameters from the \hi data. We assume only that the polarized intensity is proportional to the \hi intensity and that the emission is polarized perpendicular to the orientation of \hi filaments, as is expected from polarized dust emission. The Stokes parameters of emission with these characteristics are given by

\begin{align}
    Q_\ion{H}{i}\left(v\right) &= I\left(v\right) \sum_\theta R\left(\theta,v\right)\cos\left(2\theta\right)\label{eq:QHIv}\\ 
    U_\ion{H}{i}\left(v\right) &= I\left(v\right) \sum_\theta R\left(\theta,v\right)\sin\left(2\theta\right)\label{eq:UHIv}
    ~~~.
\end{align}
Because they are derived from the $I\left(v\right)$ data, $Q_\ion{H}{i}\left(v\right)$ and $U_\ion{H}{i}\left(v\right)$ have the same units as $I_\ion{H}{i}\left(v\right)$, K\,km\,s$^{-1}$.

The orientation $\theta_\ion{H}{i}\left(v\right)$ and ``polarization fraction'' $p_\ion{H}{i}\left(v\right)$ of the emission in each velocity channel are computed from the Stokes parameters in the usual way:

\begin{align}
    \theta_\ion{H}{i}\left(v\right) &= \frac{1}{2}\arctan\frac{U_\ion{H}{i}\left(v\right)}{Q_\ion{H}{i}\left(v\right)} \\
    p_\ion{H}{i}\left(v\right) &= \frac{\sqrt{Q_\ion{H}{i}\left(v\right)^2 + U_\ion{H}{i}\left(v\right)^2}}{I\left(v\right)}
    ~~~.
\end{align}

The additivity of the Stokes parameters permits straightforward integration over velocity channels:

\begin{figure*}
\centering
\includegraphics[width=\textwidth]{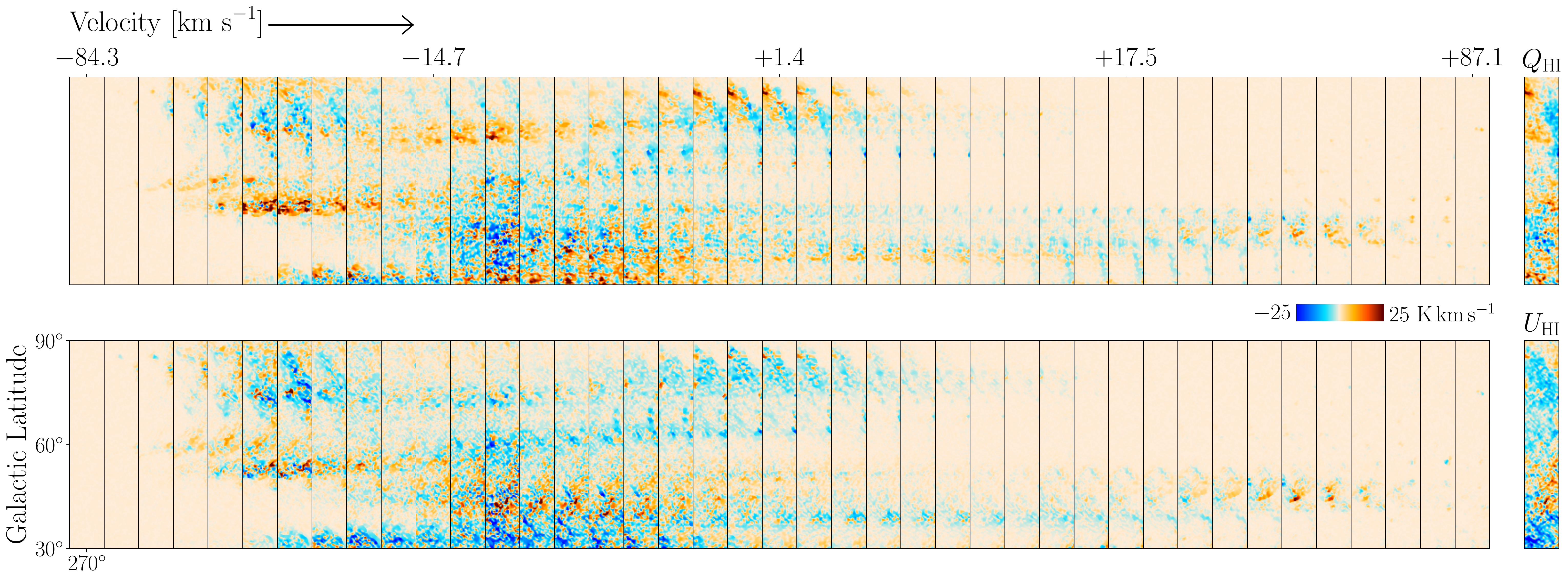}
\caption{Slices of $Q(v)$ (top) and $U(v)$ (bottom), showing our \HI-based Stokes parameters for a small region of sky in each velocity slice. Each panel shows a $10^\circ \times 60^\circ$ area of sky centered at $(l, b) = (270^\circ, 60^\circ)$. Panels correspond to the velocity channels illustrated in Figure~\ref{fig:HIbinning}. $Q(v)$ and $U(v)$ are shown at $30'$ resolution plotted on a linear scale as indicated by the colorbar. Panels on the far right show the velocity-integrated quantities $Q_{\HI}$ and $U_{\HI}$ on a linear scale from $-150\,\mathrm{K}\,\mathrm{km}\,\mathrm{s}^{-1}$ to $+150\,\mathrm{K}\,\mathrm{km}\,\mathrm{s}^{-1}$.
}\label{fig:manyvelpanel}
\end{figure*}

\begin{align}
    I_\ion{H}{i} &= \sum_v I\left(v\right)\label{eq:sumI} \\
    Q_\ion{H}{i} &= \sum_v Q_\ion{H}{i}\left(v\right) \label{eq:sumQ} \\ 
    U_\ion{H}{i} &= \sum_v U_\ion{H}{i}\left(v\right) \label{eq:sumU}
    ~~~.
\end{align}
In the magnetically coherent cloud paradigm central to this work, the sum over \hi velocities and orientations is analogous to the sum over distinct regions along the line of sight.

The orientations $\theta_\HI\left(v\right)$ and polarization fractions $p_\HI\left(v\right)$ are {\it not} additive, and so the velocity integrated orientation and polarization fraction are given by

\begin{align}
    \theta_\ion{H}{i} &= \frac{1}{2}\arctan\frac{U_\ion{H}{i}}{Q_\ion{H}{i}} \label{eq:thetaHI} \\
    p_\ion{H}{i} &= \frac{\sqrt{Q_\ion{H}{i}^2 + U_\ion{H}{i}^2}}{I_\ion{H}{i}}  \label{eq:pHI}
    ~~~.
\end{align}
Also by analogy with dust emission, the polarized intensity is defined as
\beq
P_{\HI} = p_{\HI}I_{\HI} \label{eq:PHI}
~~~.
\eeq

\begin{figure*}[h!]
\centering
\includegraphics[width=0.8\textwidth]{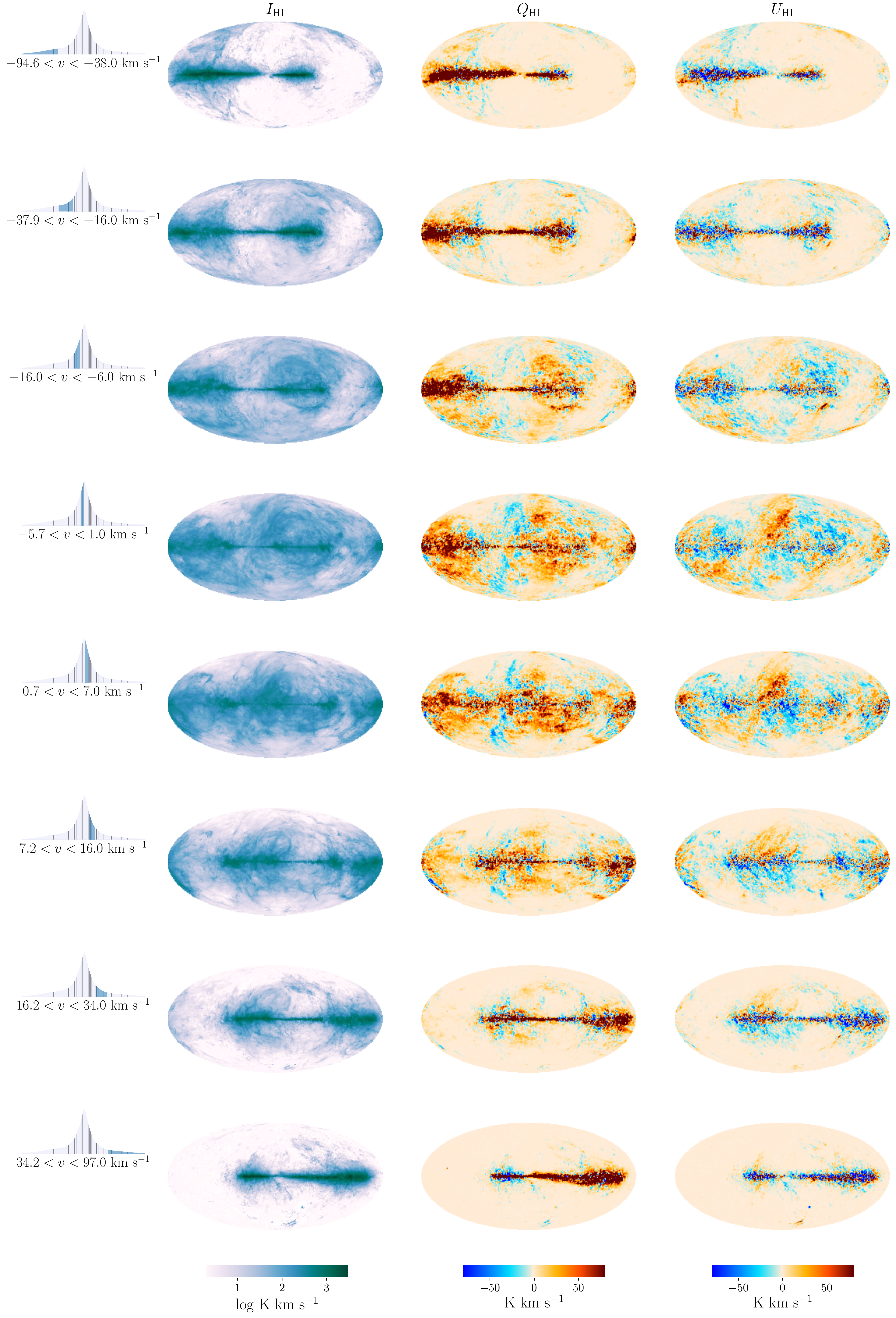}
\caption{Stokes $I_{\HI}$, $Q_{\HI}$, $U_{\HI}$ maps summed into 8 velocity bins. $I_\ion{H}{i}$ is $I(v)$ measured over the velocity range shown in the left panel, and is shown at the native \HI4PI resolution and on a logarithmic color scale to bring out fine features in the \hi distribution. $Q_\ion{H}{i}$ and $U_\ion{H}{i}$ are the result of Equations \ref{eq:sumQ} and \ref{eq:sumU} applied over the velocity range shown at $80'$. All maps are displayed in a Galactic Mollweide projection centered on the Galactic Center.}\label{fig:IQU_chunk_panelplot}
\end{figure*}

While we demonstrate that these \HI-derived quantities are effective predictors of the corresponding quantities in dust emission, a few key differences exist. First, the \hi intensity is insensitive to changes in the dust to gas ratio and dust temperature, so any variations in those quantities in the Galactic ISM will reduce the correlation with the observed dust emission.

\begin{figure*}[ht]
\centering
\includegraphics[width=\textwidth]{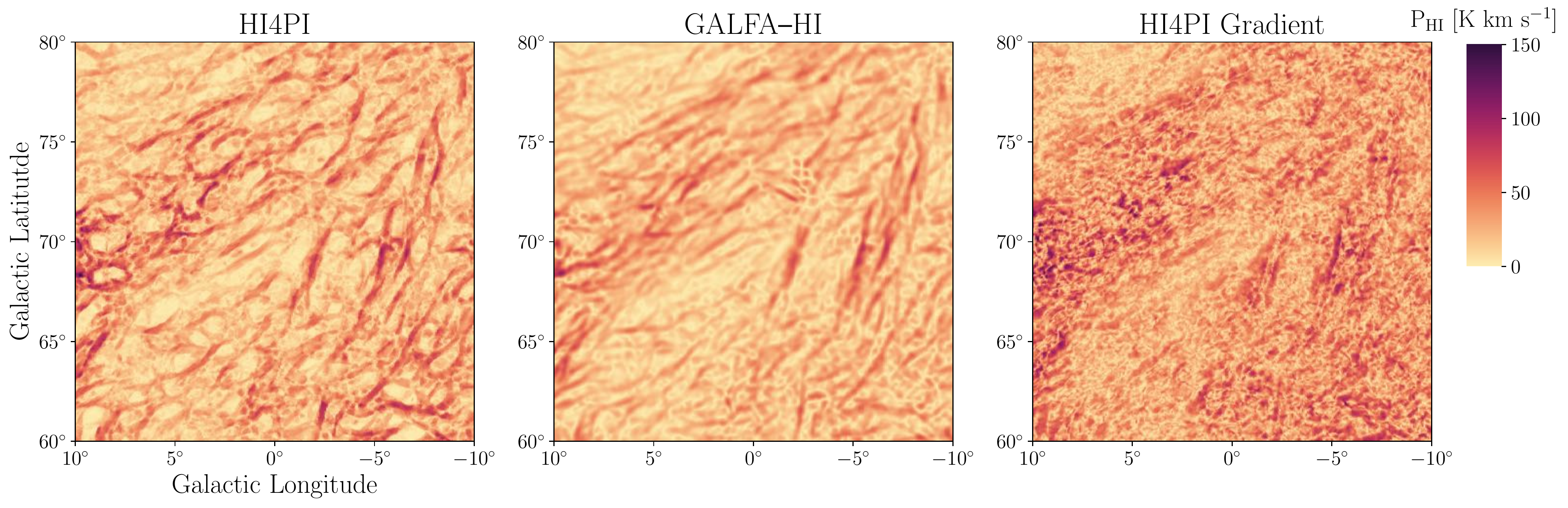}
\caption{Map of a small region of the polarized intensity $P_{\HI}$ at $16.2'$ resolution. Left and middle panels show $P_{\HI}$ from the \HI4PI and GALFA-\hi data, respectively, from Equation~\ref{eq:pHI}. Righthand panel shows $P_{\HI}$ from the \HI4PI data calculated using the spatial gradient to compute local orientations, an alternative to Equations~\ref{eq:QHIv} and \ref{eq:UHIv} discussed in Section~\ref{sec:gradient}. }\label{fig:smallareaPHI}
\end{figure*}

Second, because dust grains align with their rotation axis parallel to the local magnetic field, the polarized dust emission depends on the orientation of the Galactic magnetic field relative to the line of sight. If $\gamma$ is the angle between the magnetic field and the plane of the sky, then the polarized intensity of the dust emission is proportional to $\cos^2\gamma$ \citep{Lee:1985}. Although this angle is not directly measurable from the \hi data, it is closely related to the dispersion of polarization angles such that more dispersion is expected when the Galactic magnetic field is along the line of sight ($\gamma = \pi/2$) and less when it lies in the plane of the sky ($\gamma = 0$). This effect is captured in our maps by the sum over the distribution of orientations (Equations \ref{eq:QHIv} and \ref{eq:UHIv}). Furthermore, if linear \hi structures are elongated in the direction of the three-dimensional magnetic field, they will be less well detected when $\gamma \sim \pi/2$, as this will diminish their plane-of-sky extent. We thus expect that information about $\gamma$ is encoded in our maps implicitly.

\section{Three-dimensional I, Q, U Maps}\label{sec:3dmaps}

In this section, we present full-sky $I$, $Q$, and $U$ Stokes parameter maps as a function of velocity. These maps are the result of the equations in Section~\ref{sec:model} applied to the \HI4PI data set described in \ref{sec:data}. These maps are three-dimensional, where the third dimension is \hi velocity. Unless otherwise noted in this work, we present all Stokes parameters in the IAU Galactic linear polarization convention, in which the polarization angle is zero toward north and increases toward east \citep[see][]{Robishaw:2018}. Note that although \citet{Planck2018XII} present the polarization angle in this convention, their maps of the Stokes parameters are shown in the \texttt{COSMO} convention, which differs from the IAU standard by the sign of Stokes $U$. 

In Figure~\ref{fig:manyvelpanel} we show a small region of $Q(v)$ and $U(v)$ over the entire line of sight. The velocity-integrated quantities $Q_{\HI}$ and $U_{\HI}$ are shown in the far righthand panels. Comparing the velocity-separated maps to their integrated counterparts, one can see by eye where various features in the final map originate in velocity space. Some lines of sight intersect multiple distinct regions of magnetic coherence, while others are well described by a single structure somewhere along the line of sight. Some structures are magnetically coherent over a few to tens of \kmsa.

Our full three-dimensional Stokes parameter maps comprise 41 velocity bins each of \HI-derived $I$, $Q$, $U$ over the full sky. To display all of the data at once we partition the Stokes parameter maps into 8 velocity bins, each the sum over $\sim 5$ of the velocity bins indicated in Figure~\ref{fig:HIbinning}. These binned maps are presented in Figure~\ref{fig:IQU_chunk_panelplot}. With this full-sky view, some large-scale features are evident in the $Q(v)$ and $U(v)$ maps that have clear counterparts in the \hi emission structure. The Galactic Plane is an obvious feature of these maps, with a velocity dependence that corresponds largely to galactic rotation. Several well-known \hi shells are likewise visible in the $Q(v)$ and $U(v)$ Stokes parameter maps \citep[e.g.,][]{Heiles:1984}. 

The native resolution of these maps is nominally $16.2'$, the native resolution of the \HI4PI observations. However, we note that there is significant covariance between adjacent pixels, because the RHT measures the linearity of structures in a region of diameter $75'$ around each pixel, and so the effective resolution may be coarser.

\begin{figure*}
\centering
\includegraphics[trim=0cm 0cm 0cm 0cm,clip,width=\textwidth]{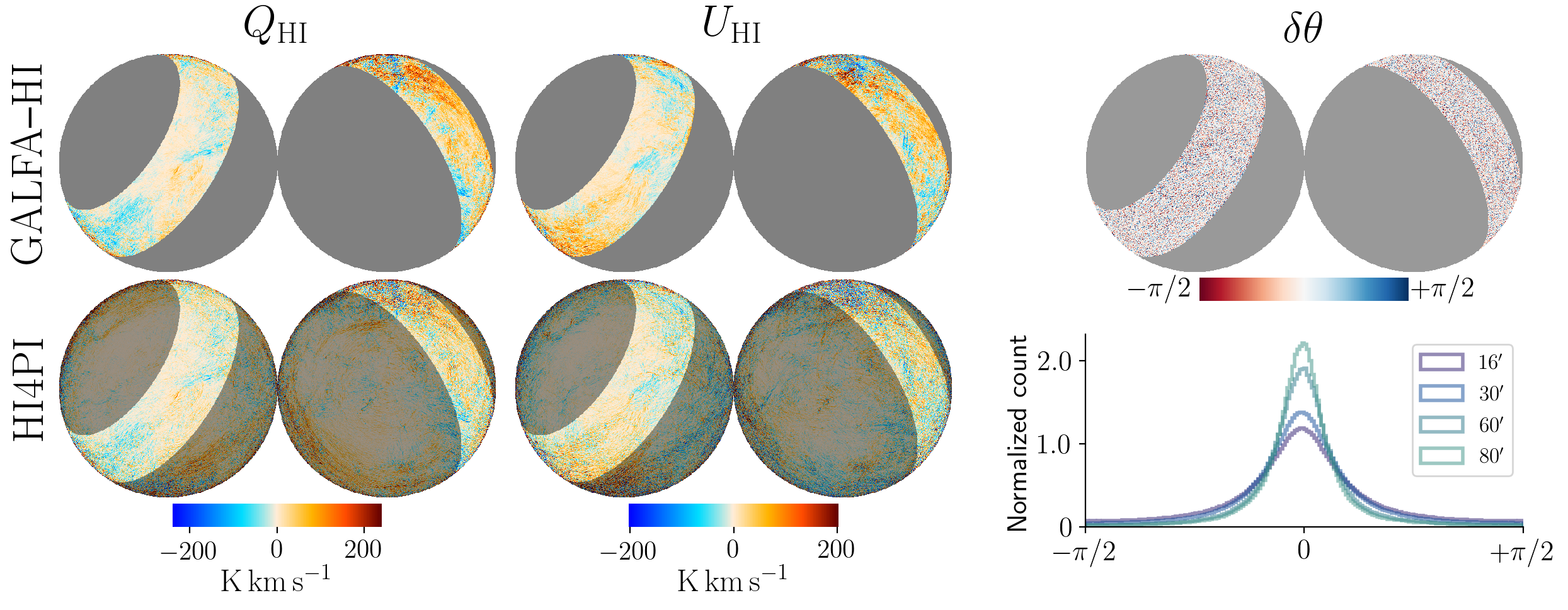}
\caption{Left: $Q_{\HI}$ and $U_{\HI}$ at $16.2'$ resolution for the GALFA-\hi (top) and \HI4PI (bottom) data. Each set of two maps shows the full sky in an orthographic projection, centered on the North Galactic Pole (left) and the South Galactic Pole (right). The GALFA-\hi footprint is shown for $1.5^\circ < decl. < 35.5^\circ$, as described in the text. The \HI4PI data cover the full celestial sphere, but the GALFA-\hi footprint is highlighted to aid in the visual comparison. Right: top panel shows the same orthographic view of $\delta \theta$ calculated between $\theta_\mathrm{HI}^\mathrm{GALFA}$ and $\theta_\mathrm{HI}^\mathrm{HI4PI}$ at $16.2'$. Bottom panel shows the histogram of this $\delta\theta$ distribution for these two quantities compared at various angular resolutions.}\label{fig:QU_GHcomp}
\end{figure*}

The small-scale structure of our maps is exemplified in Figure~\ref{fig:smallareaPHI}, which shows a small region of $P_{HI}$ centered on Galactic zenith. The filamentary geometry in the left two panels is a natural consequence of the linearity mapping we perform with the RHT, but no explicit filament boundaries have been prescribed. The idea that dusty filaments are the predominant building blocks of the polarized dust emission was recently explored explicitly by \citet{Huffenberger:2019}, who were able to reproduce statistics of the \textit{Planck} polarization data (e.g. nonunity $EE/BB$, positive $TE$ correlations) by modeling sky-projected populations of prolate spheroidal filaments. The authors use this model, based on the formalism introduced in \citet{Rotti:2019}, to explore the effect of filament properties, including aspect ratio and degree of alignment with the local magnetic field, on the polarized dust emission. Because our mapping assigns polarized intensity to the linear structures in \HI, it is interesting to compare the structure of our maps to the \citet{Huffenberger:2019} phenomenological model. Indeed, the small-scale structure of our maps is largely organized into overlapping filamentary structures that tend to be aligned with the local plane-of-sky magnetic field.

\section{Validation with GALFA-\hi}\label{sec:galfacomp}

Our all-sky maps can be compared to measurements of Galactic polarized dust emission, as we will do in the next Section. However, it is of interest to first understand the limitations of these maps. Accurate estimation of the observational uncertainty on the Stokes parameter maps is a difficult task in the absence of an \textit{a priori} model for the distribution of \hi in velocity-orientation space. We choose instead to take an empirical approach, in which we apply the equations outlined in Section~\ref{sec:model} to an independent, higher-resolution data set: the GALFA-\hi survey described in Section~\ref{sec:data}.

The GALFA-\hi data cover a different velocity range, with different velocity binning, than the \HI4PI data (Section~\ref{sec:data}). To compare the two, we compute \HI4PI-derived $I_{\HI}$, $Q_{\HI}$, and $U_{\HI}$ from a sum over a restricted velocity range: -37.3\,\kms to 40.0\,\kmsa, to hew as closely as possible to the GALFA-\hi velocity range given the \HI4PI velocity binning. We smooth the GALFA-\hi $I$, $Q$, and $U$ maps to a resolution of $16.2'$ to match the \HI4PI data.

We compare $Q_{\HI}$ and $U_{\HI}$ between these data sets, as well as $\theta_{\HI}$ from Equation~\ref{eq:thetaHI}. We find a generally good correspondence between the GALFA-\hi maps and the \HI4PI maps. Figure~\ref{fig:QU_GHcomp} shows $Q_{\HI}$ and $U_{\HI}$ maps for the two data sets, with the GALFA-\HI-based map degraded to $16.2'$ resolution. A simple linear regression of $Q_{\HI}^\mathrm{GALFA}$ vs. $Q_{\HI}^{\HI\mathrm{4PI}}$ yields a correlation coefficient $r = 0.93$ when both maps are degraded to a common $80'$ resolution. Comparing the reduced Stokes parameter $q = Q/I$ between the two maps, we find that the simple difference histogram between $q_{\HI}^{GALFA}$ and $q_{\HI}^{HI4PI}$ has $\sigma = 0.30$ for $16.2'$ maps and $\sigma = 0.18$ for $80'$ maps. As uncertainties these values significantly underestimate the fidelity of the \HI-based Stokes parameters, as differences in the RHT parameters, velocity ranges, and velocity binning between the maps will inflate this difference. 

We also compare the measurements of $\theta_{\HI}$ for each data set by computing the angular difference
\beq
\delta \theta = \frac{1}{2} \mathrm{arctan}\left[\frac{\mathrm{sin}(2\theta_{1})\mathrm{cos}(2\theta_{2}) - \mathrm{cos}(2\theta_{1})\mathrm{sin}(2\theta_{2}) }{\mathrm{cos}(2\theta_{1})\mathrm{cos}(2\theta_{2}) + \mathrm{sin}(2\theta_{1})\mathrm{sin}(2\theta_{2}) }\right]\label{eq:angdif}
\eeq
over the maps, where $\theta_1$ is $\theta_{\HI}$ computed from GALFA-\hi and $\theta_2$ is $\theta_{\HI}$ computed from \HI4PI. We show a map of $\delta \theta$ in Figure~\ref{fig:QU_GHcomp}, along with histograms of the $\delta\theta$ distribution computed at a number of resolutions. We find that a Gaussian fit to the $\delta\theta$ histogram has $\sigma = 15^\circ$ at $16.2'$ and $\sigma = 8.9^\circ$ at $80'$. We can consider this an empirical value for the uncertainty in our maps, although the different velocity binning considered in the two maps likely means that this is an overestimate. The map distribution of $\delta\theta$ does not show obvious structure.

We discuss here the possible sources of statistical or systematic error that enter our calculations of $Q_{\HI}$ and $U_{\HI}$. Small-scale radiometer noise in the \HI4PI or GALFA-\hi maps is unlikely to contribute much to the measurement of \hi orientation because of the relatively large window used to derive $R(v, \theta)$, but does exist in $I(v)$. Telescope scan artifacts can introduce linear systematics that will affect $Q_{\HI}$ and $U_{\HI}$, and such artifacts are of course uncorrelated between \HI4PI and GALFA-\HI. The RHT was applied to the GALFA-\hi data in an Equatorial coordinate projection, and the resulting $I_{\HI}$, $Q_{\HI}$, $U_{\HI}$ maps were transformed to Galactic coordinates by a simple rotation. This rotation was applied at the pixel level at $N_{\rm side} = 2048$ for the $4'$ GALFA-\hi maps, well below the resolution of the \HI4PI data. Comparisons between the RHT output computed in a given projection vs. RHT output computed in one projection and transformed to another revealed no obvious systematic differences between those two approaches.

Crucial to the interpretation of the difference between the GALFA-\hi and \HI4PI maps is the fact that computing $R(v, \theta)$ on maps at a given resolution and smoothing those maps are not commutative procedures. Linear structures that are prominent at high resolution may not be detected with as sharp an $R(\theta)$ distribution at lower resolution, or if they are significantly washed out, may slip below the detection threshold entirely. These effects are discussed at length in \citet{Clark:2014}. Figure~\ref{fig:smallareaPHI} shows a small region of $P_{\HI}$ for both the \HI4PI and GALFA-\hi data, where the GALFA-\HI-based measurement has been degraded to the $16.2'$ resolution of the \HI4PI data. While the maps show a similar morphology, the visual differences between them are likely attributable to better-resolved, thin linear structures in the GALFA-\hi data. The angular size of the structures that are most predictive of the magnetic field orientation is a fundamental physical limitation on the resolution needed to probe the three-dimensional magnetic field structure with \hi data. 

\begin{figure*}[h!]
\centering
\includegraphics[width=0.9\textwidth]{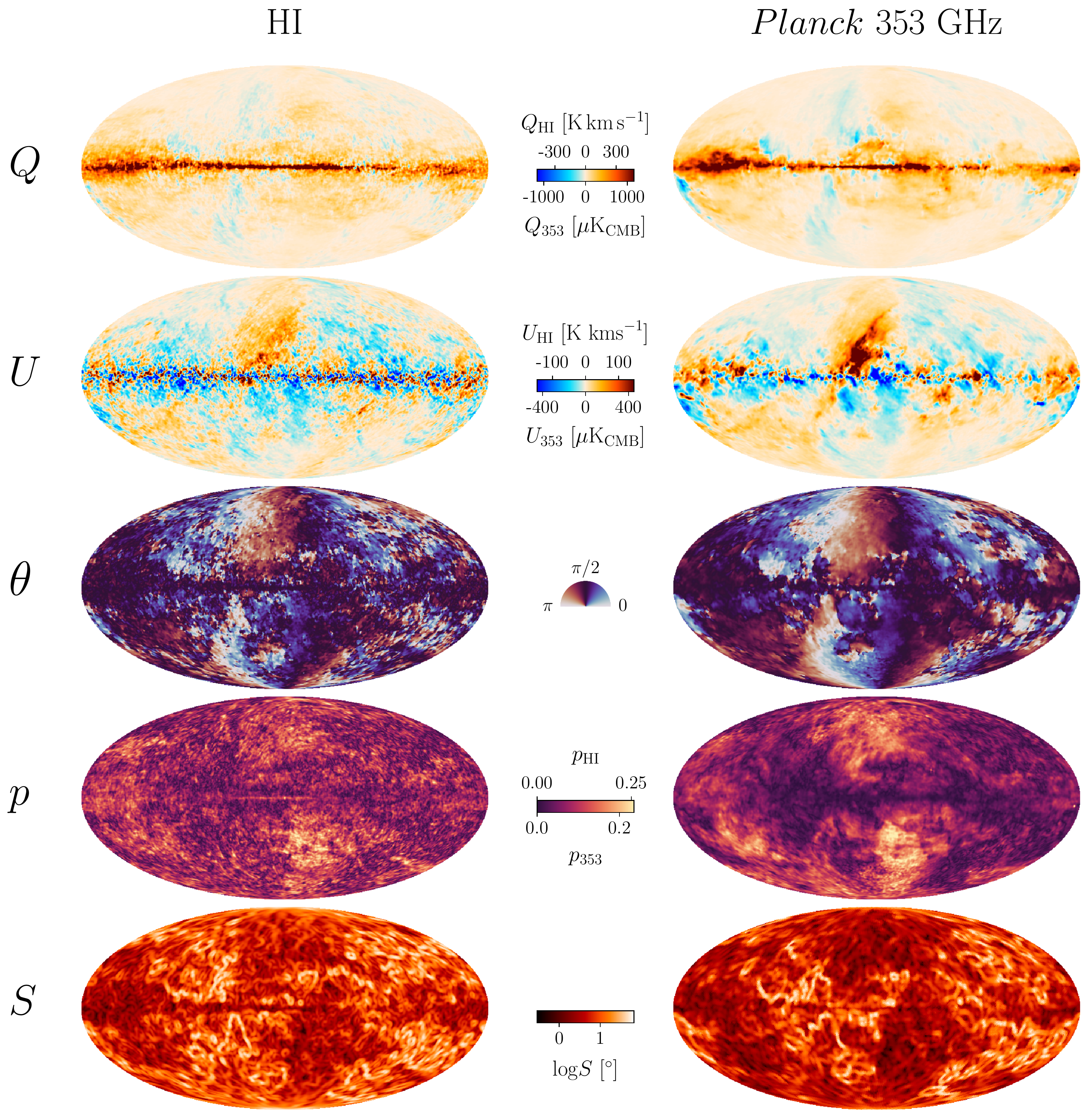}
\caption{\hi maps (left) compared to \textit{Planck} 353\,GHz data (right). From top to bottom, the maps are Stokes $Q$, Stokes $U$, the plane-of-sky magnetic field orientation $\theta$, the polarization fraction $p$, and the polarization angle dispersion function $S$. All maps are shown at a resolution of $80'$ except for $S$, which is at $160'$. The Stokes parameter maps are shown in two units, indicated on the top and bottom of a shared color bar. $Q_{\HI}$ and $U_{\HI}$ are in $\mathrm{K}\,\mathrm{km}\,\mathrm{s}^{-1}$, while $Q_{353}$ and $U_{353}$ are in $\mu\mathrm{K}_\mathrm{CMB}$. The magnetic field orientation angles $\theta_{\HI}$ and $\theta_{353}$ are plotted on their shared half-polar range $[0, \pi)$. The polarization fractions $p_{\HI}$ and $p_{353}$ are plotted on a linear scale between 0 and their respective $99.9^{th}$ percentile values. $\mathcal{S}_{\HI}$ and $\mathcal{S}_{353}$ are plotted on a shared logarithmic scale. }\label{fig:sixpanel_planck_HI}
\end{figure*}

\begin{figure*}
\centering
\includegraphics[width=0.9\textwidth]{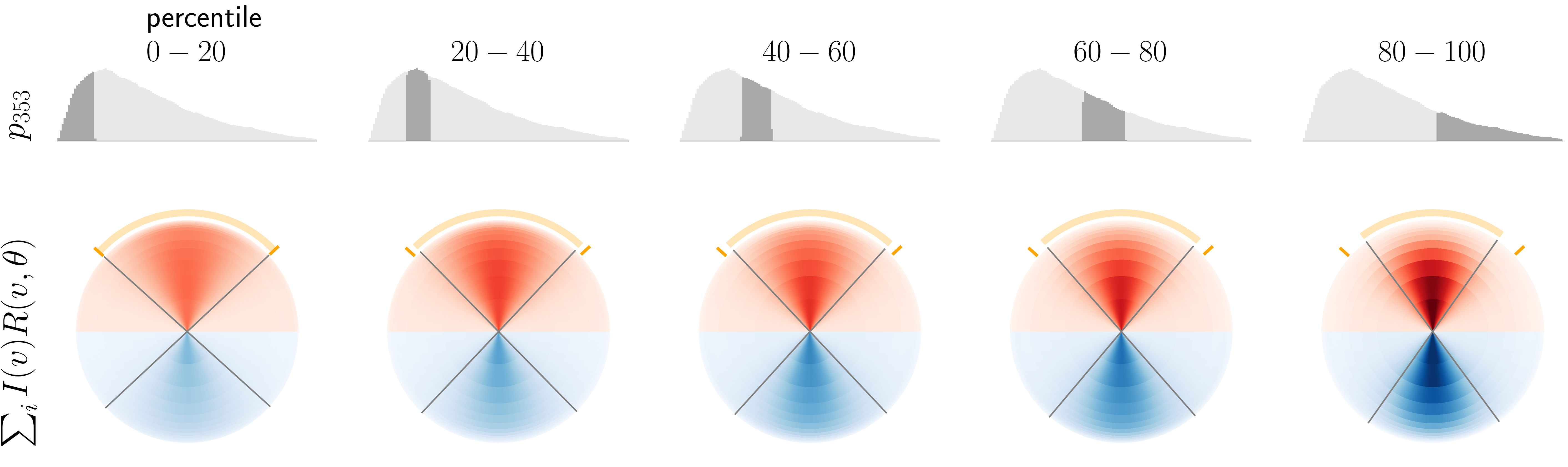}
\caption{All of the $I(v)R(v, \theta)$ data integrated over regions of sky partitioned into quintiles of $p_{353}$, the \textit{Planck} 353\,GHz polarization fraction (top). The \hi data associated with each pixel are plotted relative to their respective mean orientation $\theta_{\HI}$ (bottom). The summed $I(v)R(v, \theta)$ in each quintile is normalized, and all quintiles are plotted on the same color scale. Darker red and blue thus correspond to higher peaks in the normalized $I(v)R(v, \theta)$. Orange semicircles and gray lines denote the FWHM of the mean of the $I(v)R(v, \theta)$ distribution along the velocity axis: that is, the $I$-weighted mean $R(\theta)$. This is $\mathrm{FWHM}=95^\circ$ for the data with $p_{353}$ in the first quintile, and that FWHM is indicated by the orange lines on each percentile plot for the sake of comparison. Values for the remaining quintiles are FWHM = $88^\circ$, $86^\circ$, $81^\circ$, $72^\circ$, respectively. As the \textit{Planck} polarization fraction increases, the typical \hi distribution in velocity-orientation space becomes more collimated and peaked about its mean. }\label{fig:quintiles}
\end{figure*}

\section{Comparison to \textit{Planck} polarized dust emission}\label{sec:planckcomp}
Our maps predict the structure of polarization in three dimensions, without prescribing specific dust properties or modeling the dust SED. We compare our \HI-based polarization maps to \textit{Planck} observations of polarized dust emission at 353\,GHz described in Section~\ref{sec:data}, the most sensitive full-sky dust polarization measurements available. As the \hi model maps are constructed entirely from the spatial and velocity distribution of the \HI, comparing this model to the \textit{Planck} 353\,GHz measurements can elucidate how the polarized dust emission is related to the distribution of the three-dimensional ISM. It is of interest to examine regions of the sky where the \hi maps are strongly correlated with the \textit{Planck} measurements, as well as regions where the two maps diverge.

Figure~\ref{fig:sixpanel_planck_HI} shows five polarization quantities for our \HI-based maps and for \textit{Planck}: $Q$, $U$, the magnetic field orientation angle, the polarization fraction, and the polarization angle dispersion function, discussed further below. The overall visual correspondence between the purely \HI-based maps and the 353\,GHz observations is striking. The large-scale gradient in the polarization angle near the poles in both maps is a manifestation of the Galactic Stokes coordinate system: a constant magnetic field orientation projected onto the plane of the sky will have a longitude-variable $\theta$ in a reference frame defined in relation to the North Galactic Pole. The similarity between the \HI-based map and the polarized dust emission is however not a projection effect, and indeed persists in coordinate systems that remove this large-scale gradient. In this section we examine some correlations between these quantities in detail.

\subsection{Polarization fraction}\label{sec:polfrac}

We first demonstrate in Figure~\ref{fig:quintiles} that the velocity-orientation space mapping is predictive of the fractional polarization of each sightline, by showing that the distribution of $R(v, \theta)$ is sensitive to the measured \textit{Planck} 353\,GHz polarization fraction, $p_{353}$. We bin the maps into quintiles of $p_{353}$, and observe that the $R(v, \theta)$ distribution becomes more narrowly peaked about its mean for sightlines with higher polarization fractions. This confirms our intuition that lines of sight with \hi that is less coherent in velocity-orientation space will be more depolarized, as emission from dusty regions with different magnetic field orientations adds vectorially.

\begin{figure}
\centering
\includegraphics[width=\columnwidth]{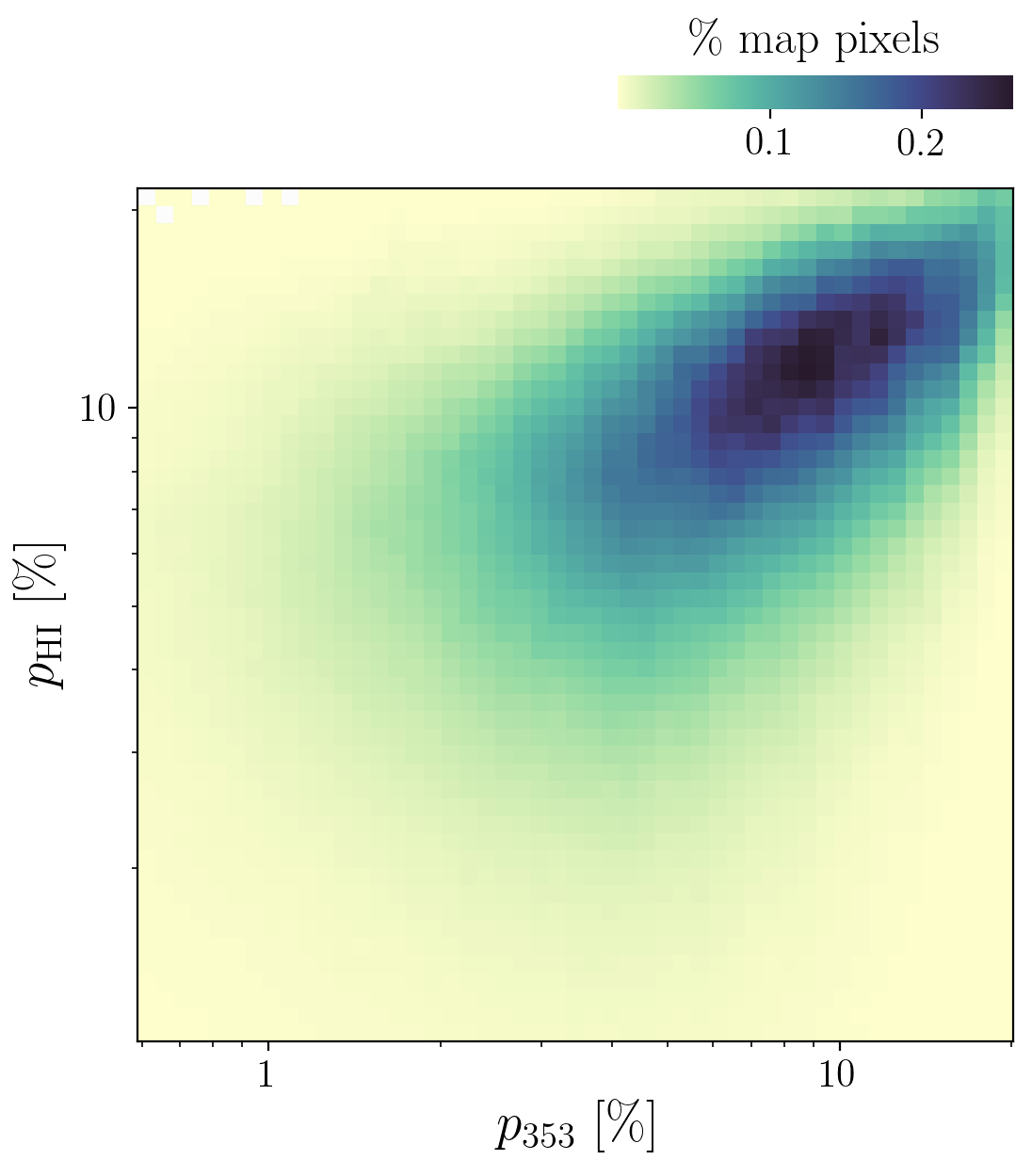}
\caption{Two-dimensional histogram of $p_{353}$ vs. $p_{\HI}$, both at $80'$ resolution. Color map indicates histogram count as a fraction of the total number of map pixels. Histogram bins are spaced logarithmically between the 1$^{st}$ and 99$^{th}$ percentile values of each data set.} \label{fig:pvsp2dhist}
\end{figure}

Figure~\ref{fig:sixpanel_planck_HI} includes all-sky maps of both $p_{353}$ and $p_{\HI}$, and we show a two-dimensional histogram of the two polarization fractions in Figure~\ref{fig:pvsp2dhist}. The two quantities are strongly linearly correlated, with appreciable scatter. A simple linear regression of $p_{353}$ vs. $p_{\HI}$ over the whole sky yields a correlation coefficient $r \sim 0.6$. As $p_{\HI}$ is derived solely from the morphology of $I(v)$, this correlation indicates that much of the variation in $p_{353}$ arises purely from the geometry of the magnetic field rather than, e.g., variable grain alignment, in good agreement with the findings in \citet{Planck2018XII} and \citet{Clark:2018}.

The maximum value of the observed polarization fraction is an important constraint on the intrinsic polarizing efficiency of dust grains \citep{Draine:2009,Guillet:2018}. \citet{Planck2018XII} reported a maximum polarization fraction $p_{353}^{\rm max} \sim 0.22$. This implies that dust grain populations must emit thermal radiation that is intrinsically {\it at least} $\sim 22\%$ polarized, before effects like magnetic field tangling induce depolarization. By construction, $p_{\HI}$ is sensitive only to these geometrical effects and not to the intrinsic polarization efficiency of dust. One might therefore expect that the most coherent sightlines have $p_{\HI} \sim 1$, but Figures~\ref{fig:sixpanel_planck_HI} and \ref{fig:pvsp2dhist} show that the range of $p_{\HI}$ is similar to $p_{353}$: the $99.9^{th}$ percentile value of $p_{\HI}$ is $\sim 0.26$ at $80'$ resolution. 

If $p_{\HI}$ were a numerically accurate predictor of the effects of geometric depolarization, this would imply that dust emission must have an intrinsic polarizing efficiency of nearly unity. However, the alignment of \hi gas with magnetic field lines is imperfect, and so we generically expect the \hi orientations to be more disordered than the true magnetic field. As a consequence, $p_{\HI}$ is an overestimate of the geometric depolarization present in the dust maps. Another reason $p_{\HI}$ may overestimate the amount of depolarization is that the denominator of Equation~\ref{eq:pHI} includes {\it all} of the \hi emission along the line of sight, while $Q_{\HI}$ and $U_{\HI}$ are weighted only by the emission in linear \hi features. Since there are some velocity bins in which $R(v,\theta)$ is zero for all $\theta$, these bins contribute to the total but not polarized intensity (Section~\ref{sec:coherentclouds}). This likely undercounts the fraction of \hi emission that is correlated with polarized dust emission, as discussed further in Section~\ref{sec:discussion}. The numerical proximity of $p_{\HI}^{\rm max}$ to $p_{353}^{\rm max}$ appears therefore to be a coincidence of the overestimation of depolarization compensating the absence of an intrinsic polarization fraction in the model.

\subsection{The magnetic field orientation}

We can compare the estimate for the magnetic field orientation between two maps by computing the angular difference (Equation~\ref{eq:angdif}) in each pixel, where $\theta_1$ in this case is $\theta_{\HI}$ computed from the relevant velocity range, and $\theta_2$ is $\theta_{353}$. We collapse this $\delta\theta$ distribution to a point statistic for the mean degree of alignment, 

\beq\label{eq:xi}
\mathrm{\xi} = \left< \mathrm{cos} \phi \right>,
\eeq
where

\beq
\phi = 2 \delta \theta.
\eeq 

When $\theta_{\HI} = \theta_{353}$, $\phi = 0$, and when $\theta_{\HI}$ and $\theta_{353}$ are orthogonal to one another, $\phi = \pi$. The metric $\xi$ is therefore defined on $[-1, 1]$, such that two perfectly aligned distributions will have $\xi = 1$, two anti-aligned distributions will have $\xi = -1$, and two distributions with no statistical alignment will have $\xi = 0$. This metric differs from the ``projected Rayleigh statistic", \citep[PRS;][]{Jow:2018}, by a factor of $\sqrt{2 N}$, where $N$ is the number of samples (pixels). That is,

\beq
\mathrm{PRS} = \sqrt{\frac{2}{N}} \sum_i \mathrm{cos}\, \phi_i,
\eeq
and the uncertainty can be estimated as 

\beq
\sigma^2_{PRS} = \frac{2}{N} \left[\sum_i \mathrm{cos}^2\, \phi_i - (PRS)^2 \right].
\eeq

The PRS is equivalent to the modified Rayleigh test for uniformity proposed by \citet{Durand:1958}, for the specific case where the mean angle of the distribution is $\phi=0$. We define $\xi$ for its convenient range, but will also report the PRS where appropriate. For the full-sky $\theta_{\HI}$ and $\theta_{353}$ maps shown in Figure~\ref{fig:sixpanel_planck_HI}, $\xi = 0.71$. For the $80'$ maps pixelated on an $N_{\rm side}=128$ HEALPix grid, PRS=446 and $\sigma_\mathrm{PRS}=0.5$. 

\subsection{The polarization angle dispersion function}

\begin{figure}
\centering
\includegraphics[width=\columnwidth]{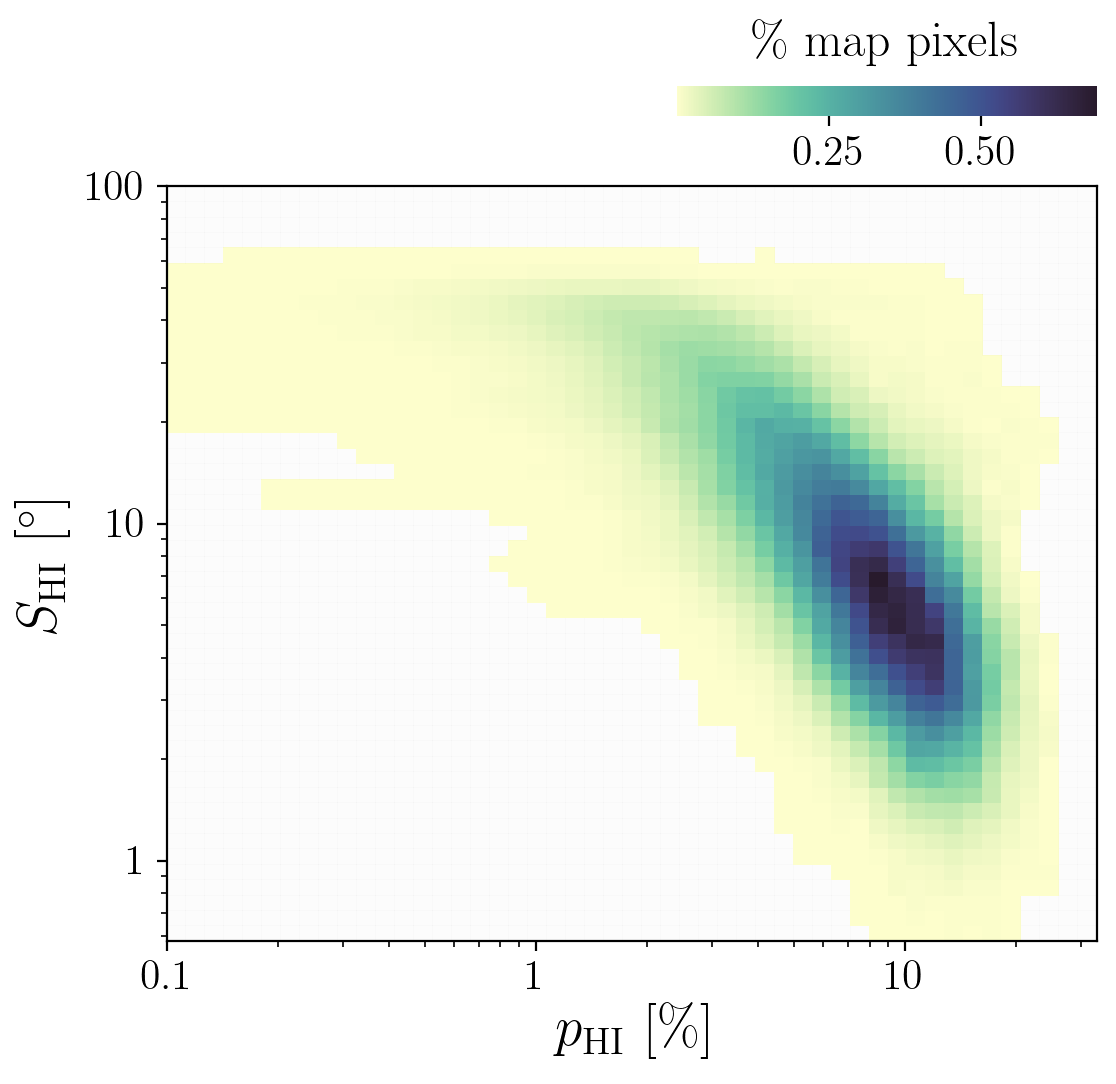}
\caption{Two-dimensional histogram of $p_{\HI}$ vs. $\mathcal{S}_{\HI}$, both at $160'$ resolution. Bins and colormap are as in Figure~\ref{fig:pvsp2dhist}.}
\label{fig:pvsS2dhist}
\end{figure}

\begin{figure}
\centering
\includegraphics[width=\columnwidth]{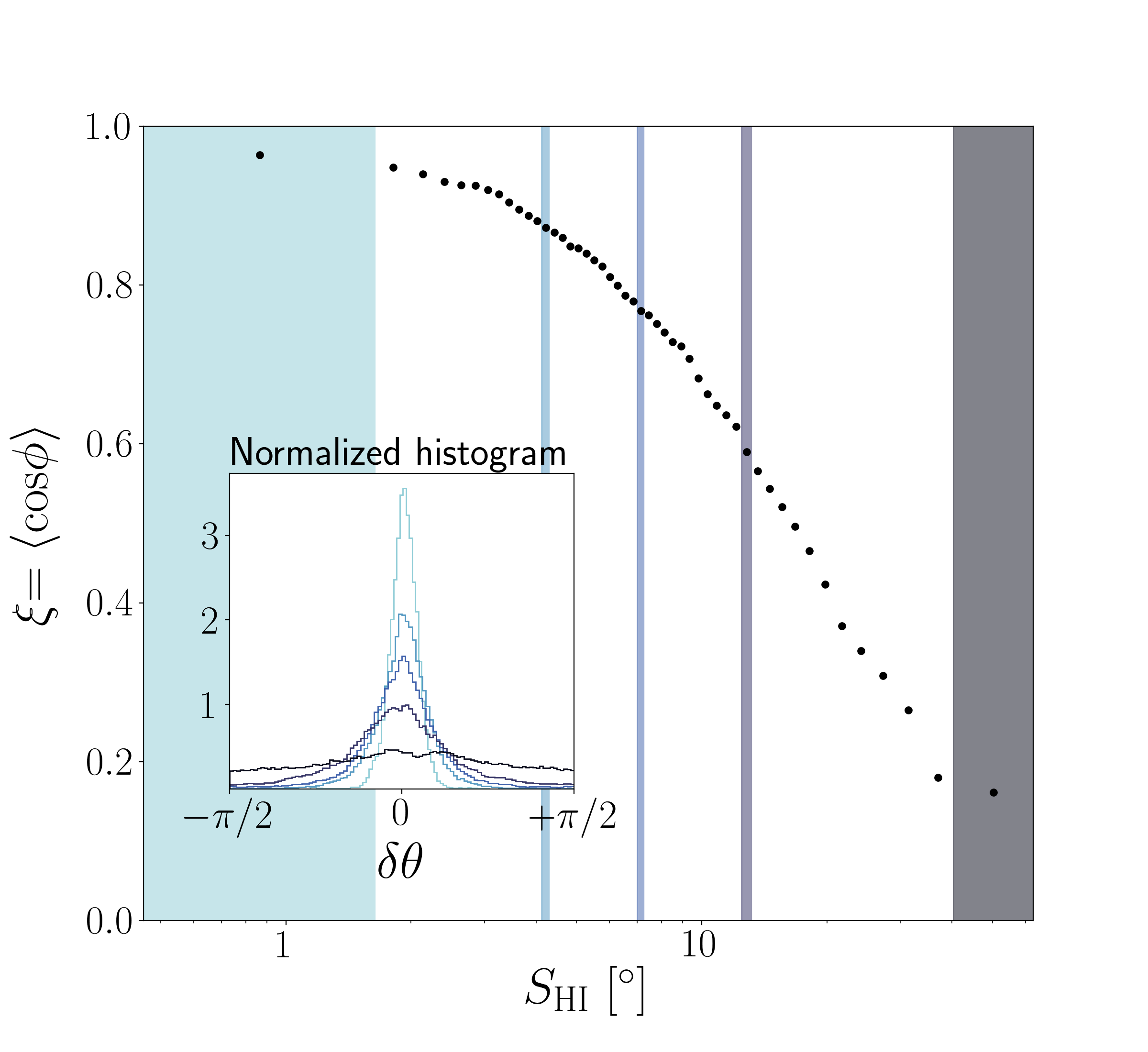}
\caption{The mean degree of alignment ($\xi$, Equation~\ref{eq:xi}) between $\theta_{\HI}$ and $\theta_{353}$ as a function of $\mathcal{S}_{\HI}$. Plot background is color coded by 5 bins of $\mathcal{S}_{\HI}$, and the inset shows histograms of $\delta \theta$ in colors corresponding to those same $\mathcal{S}_{\HI}$ bins. Bins are spaced evenly in percentiles of $\mathrm{log}\,\mathcal{S}_{\HI}$, such that each bin contains approximately the same number of pixels. The \HI-based magnetic field orientation is significantly better aligned with the \textit{Planck} magnetic field orientation in regions of low polarization angle dispersion.} \label{fig:deltatheta_S}
\end{figure}

One measure of the degree of order in the plane-of-sky magnetic field orientation is the polarization angle dispersion function, $\mathcal{S}$ \citep{PlanckXIX}. Following the notation used by the \textit{Planck} collaboration, we define

\beq
S(\mathbf{r}, \delta) = \sqrt{ \frac{1}{N} \sum_{i=1}^{N} \left[\psi (\mathbf{r} + \delta_i) - \psi(\mathbf{r})  \right]^2 },
\eeq
where the sum is over pixels located within an annulus of inner radius = $\delta/2$ and outer radius = $3\delta/2$. In practice, $\mathcal{S}$ is computed in terms of the Stokes parameters $Q$ and $U$, and in order to avoid a spuriously high $\mathcal{S}$ near the poles from the polarization reference frame, we calculate $\mathcal{S}$ such that each annulus sits at the equator of its reference frame. \citet{Planck2018XII} found that polarization systematics bias the measurement of $\mathcal{S}$ at high Galactic latitudes unless $\mathcal{S}$ is computed at a resolution of $160'$ and lag $\delta = 80'$, so we adopt this resolution and lag for both $\mathcal{S}_{\HI}$ and $\mathcal{S}_{353}$. 

Figure~\ref{fig:sixpanel_planck_HI} includes all-sky maps of $\mathcal{S}_{\HI}$ and $\mathcal{S}_{353}$. The $\mathcal{S}_{\HI}$ parameter derived from $Q_{\HI}$ and $U_{\HI}$ is broadly similar to $\mathcal{S}_{353}$ derived from the \textit{Planck} observations, particularly on large angular scales. In Figure~\ref{fig:pvsS2dhist} we examine the relationship between $p_{\HI}$ and $\mathcal{S}_{\HI}$. We find a similar trend to the relationship that \citet{Planck2018XII} reported for $p_{353}$ and $\mathcal{S}_{353}$: there exists a negative correlation between $\mathcal{S}$ and $p$ \citep[see also][]{Fissel:2016}. When the polarization angles used to compute a value of $\mathcal{S}$ are completely randomly oriented, $\mathcal{S} = \pi/\sqrt{12}$ \citep{PlanckXIX,Planck2018XII}. This is why the values of $\mathcal{S}_{\HI}$ asymptote near $52^\circ$. 

We also examine the correlation between the magnetic field orientation of our maps, $\theta_{\HI}$, and the \textit{Planck} $\theta_{353}$ measurement, as a function of the polarization angle dispersion function (Figure~\ref{fig:deltatheta_S}). We compute Equation~\ref{eq:xi} for data binned by $\mathcal{S}_{\HI}$ value. We find a striking inverse correlation between $\mathcal{S}_{\HI}$ and $\xi$, indicating that in regions of sky where the plane-of-sky magnetic field orientation is relatively ordered within a $160'$ beam (low $\mathcal{S}_{\HI}$), the magnetic field orientation in our \HI-based maps is well aligned with the magnetic field orientation inferred from the 353\,GHz dust measurements (high $\xi$). Conversely, regions of high polarization angle dispersion see a significant decrease in the statistical alignment of the two sets of angles. In all bins $\theta_{353}$ and $\theta_{\HI}$ are significantly aligned: $PRS = 691, \sigma_{PRS} = 0.1$ in the lowest $\mathcal{S}_{\HI}$ bin, and $PRS = 116, \sigma_{PRS} = 1$ in the highest $\mathcal{S}_{\HI}$ bin. As discussed further in Section~\ref{sec:HIgamma}, this is qualitatively consistent with the hypothesis that at $160'$, regions of low $\mathcal{S}_{\HI}$ are preferentially regions where the mean magnetic field orientation lies in the plane of the sky.

\subsection{Comparison of integrated quantities as a function of \hi velocity range}

The polarized emission at 353\,GHz is optically thin, so the \textit{Planck} observations trace dust emission integrated over the entire line of sight. With our velocity-resolved Stokes parameters, we can examine the correlation between the \textit{Planck} measurements and the \HI-based maps as a function of the \hi velocity range considered, by changing the bounds of the sums in Equations~\ref{eq:sumQ} and \ref{eq:sumU}. The data are summed over an ever-widening velocity range that is always centered at the same velocity bin ($v \sim 0$\,\kmsa).

\begin{figure}
\centering
\includegraphics[width=\columnwidth]{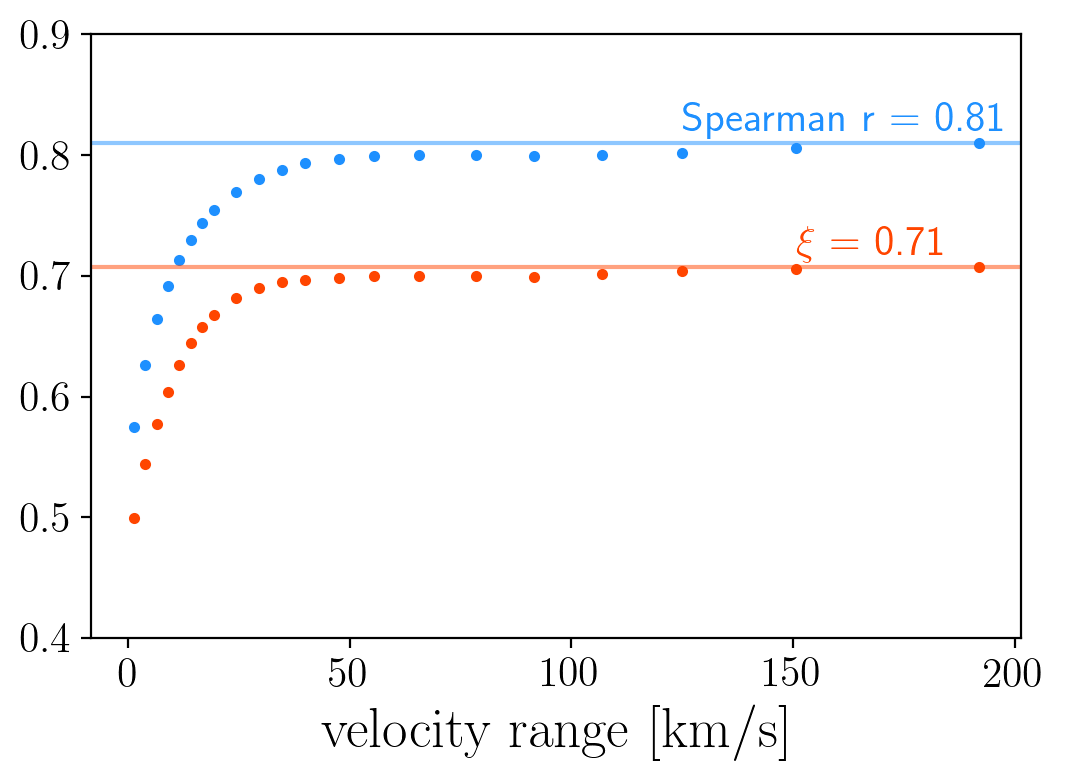}
\caption{Correlations between the \HI-based maps and the \textit{Planck} 353\,GHz polarization measurements as a function of the \hi velocity range used in Equations~\ref{eq:sumQ} and \ref{eq:sumU}. Blue dots show the Spearman rank correlation coefficient between $Q_{\HI}$ and $Q_{353}$. Red dots show the mean alignment, $\xi$ (Equation~\ref{eq:xi}), between $\theta_{\HI}$ and $\theta_{353}$. Lines show the values of each of these quantities for the full \hi velocity range considered in this work, $-93.9 < v < 96.7$\,\kmsa.  }\label{fig:corr_vs_vel}
\end{figure}

We examine the Spearman correlation coefficient between $Q_{\HI}$ and $Q_{353}$ as a function of the velocity range used in this sum. We find a monotonic increase in the correlation, with a steep dependence on the velocity range until about 30\,\kmsa, followed by a gradual improvement of the correlation out to the full velocity range. We repeat this experiment but compare $\theta_{\HI}$ and $\theta_{353}$ by computing Equation~\ref{eq:xi}. We find the same qualitative dependence on the velocity range for \ref{eq:xi} as we do for the $Q_{\HI}$ - $Q_{353}$ correlation. Figure~\ref{fig:corr_vs_vel} shows these correlations over the full sky, but we find a similar trend when restricting the latitude range to $|b| > 30^\circ$ or $|b| > 60^\circ$.

There is on average less Galactic \hi emission far from the 21-cm rest wavelength, and the distribution of \hi intensity along the line of sight is incorporated into $Q_{\HI}(v)$ and $U_{\HI}(v)$ via the $I(v)$ weighting in Equations~\ref{eq:QHIv} and \ref{eq:UHIv}. There do exist \HI-bright structures at large absolute velocities, notably high-velocity clouds \citep[HVCs;][]{Putman:2012} and nearby galaxies, including the Magellanic Clouds. In general we expect contamination from these extragalactic objects to decrease the correlation between our \HI-based maps and the \textit{Planck} data, as the component-separated 353\,GHz maps in principle trace only Galactic dust. This is particularly true in the case of HVCs, which are \hi rich but dust deficient \citep{Wakker:1986, Planck2011XXIV, Lenz:2017}. We have purposefully restricted the velocity range of the \hi gas that we consider in this work to $-93.9 < v < 96.7$\,\kms in order to avoid the bulk of this high-velocity emission. This is consistent with the velocity range identified by \citet{Lenz:2017} for which \hi column density and dust reddening are most strongly correlated. The monotonic increase of the quantities in Figure~\ref{fig:corr_vs_vel} suggests that our velocity cutoff is sufficient to avoid significant decorrelation from extragalactic gas, but this statement is only valid as a global average, and individual sightlines may contain significant emission associated with non-Galactic gas. It is straightforward to restrict the velocity range of our 3D maps for studies of such regions.

\subsection{Cross-power spectra}
\label{subsec:cross_spectra}

\begin{figure}
\centering
\includegraphics[width=\columnwidth]{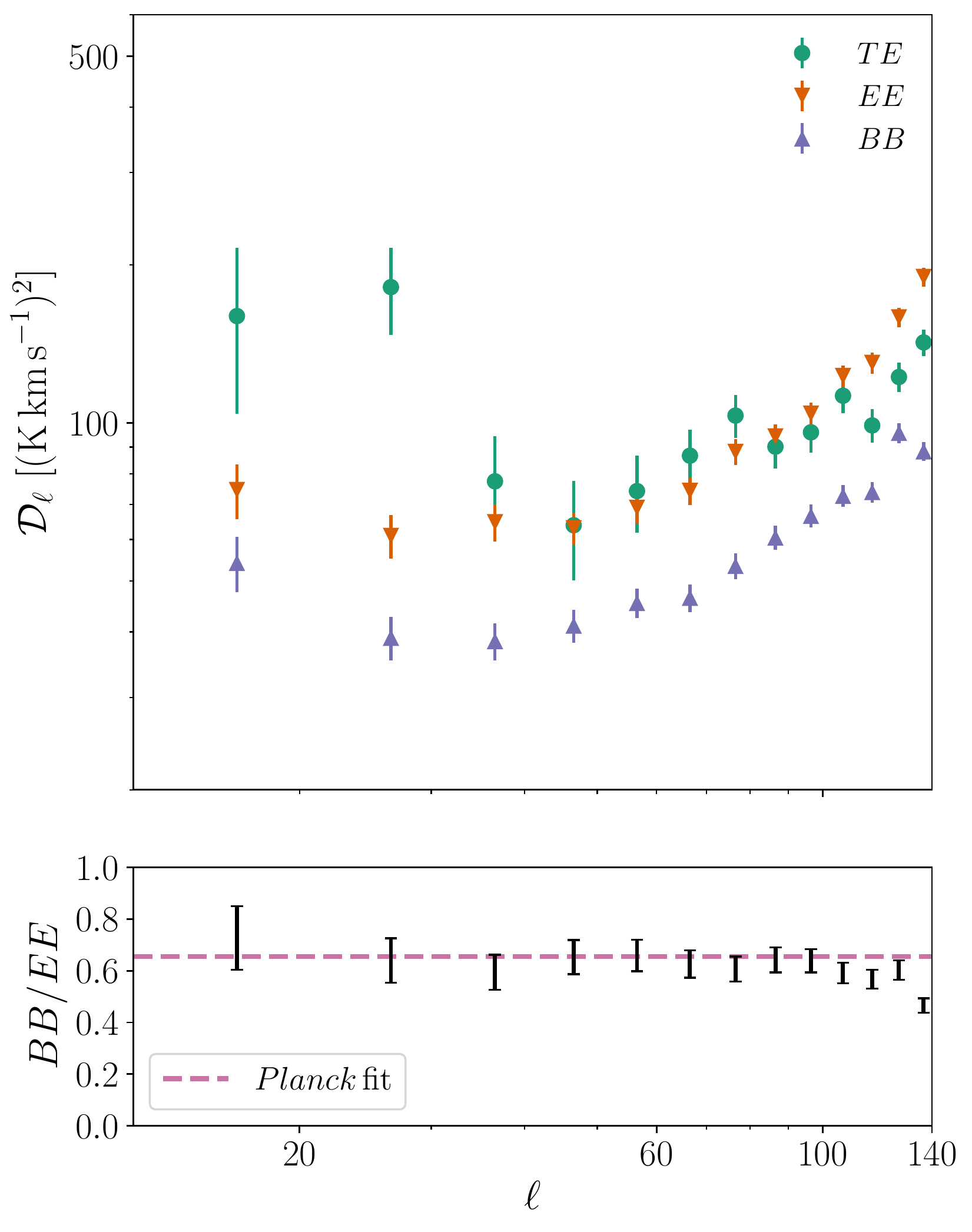}
\caption{The $TE$, $EE$, and $BB$ cross-power spectra $\mathcal{D}_\ell$ computed from $I_{\HI}$, $Q_{\HI}$, $U_{\HI}$, as described in Section~\ref{subsec:cross_spectra} (top). Error bars include sample variance only (Equation~\ref{eq:d_ell_err}). The corresponding $BB/EE$ from the \HI-based maps is shown in the bottom panel. Dashed pink line indicates the $BB/EE$ ratio calculated from power-law fits to the \textit{Planck} 353\,GHz half-mission data in the same region of sky. The $E/B$ asymmetry observed in the dust emission is reproduced by the \hi maps.} \label{fig:cross_spectra}
\end{figure}

\begin{figure}
\centering
\includegraphics[width=\columnwidth]{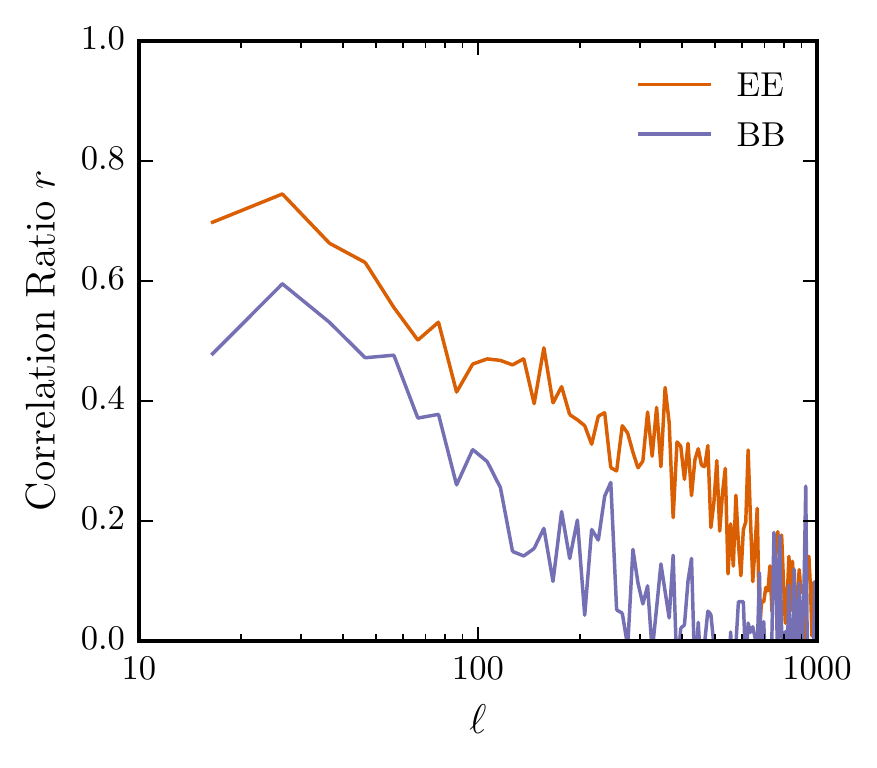}
\caption{The correlation ratios $r^{EE}$ and $r^{BB}$ between the \hi maps and the {\it Planck} 353\,GHz dust emission maps. The correlation is strongest on large scales and declines toward small scales.} \label{fig:r_planck}
\end{figure}

In the previous sections we demonstrated that the \HI-derived Stokes maps reproduce many key features of the 353\,GHz polarized dust emission at the map level. The interest in polarized dust emission as a foreground for CMB science has necessitated characterization of polarized dust emission at the power spectrum level as well, and so in this section we examine the properties of the $TE$, $EE$, and $BB$ spectra.

We compute the power spectra $\mathcal{D}_\ell^{XY} \equiv \ell(\ell+1)C_\ell/2\pi$ for our \HI-derived Stokes maps, where $XY$ are $TE$, $EE$, and $BB$. $C_\ell$ represents the pure-$E$ and pure-$B$ pseudo-$C_\ell$ estimator, computed to eliminate $E$-$B$ mixing on our cut-sky maps \citep{Smith:2006}. All cross spectra are computed with the \texttt{NaMaster} software \citep{Alonso:2019}. We restrict our analysis to $|b| > 30^\circ$, but find that our results are stable to more aggressive masking. To facilitate comparison with the {\it Planck} 353\,GHz power spectra, we mask all pixels in the {\it Planck} half-mission $I$, $Q$, and $U$ maps that have been set to \texttt{bad\_value}. In addition, we mask all pixels for which $\sum_v\sum_\theta R(v, \theta) = 0$. This mask is apodized with the analytic $C^2$ method of \citet{Grain:2009}, with an apodization scale of $5^\circ$. These cuts result in a sky fraction $f_{\rm sky} = 41\%$. 

In Figure~\ref{fig:cross_spectra}, we present the three cross spectra binned with $\Delta\ell = 10$. The error bars are computed assuming simple Gaussian sample variance and no instrumental noise, i.e.,

\begin{equation}
    \label{eq:d_ell_err}
    \sigma_{\mathcal{D}^{XY}_\ell} = \sqrt{\frac{\mathcal{D}^{XX}_\ell \mathcal{D}^{YY}_\ell + \left(\mathcal{D}^{XY}_\ell\right)^2}{f_{\rm sky}\left(2\ell+1\right)\Delta\ell}}
    ~~~,
\end{equation}
and so are somewhat underestimated.

On the largest scales ($\ell \lesssim 50$), the cross spectra have a shallow negative slope. At higher $\ell$, and unlike what is observed in the dust emission \citep{Planck2018XI}, the cross spectra begin to rise with increasing $\ell$, suggesting an excess of small-scale power in the \hi maps relative to the dust emission. This behavior is perhaps unsurprising since all polarized emission in this model is explicitly attributed to small-scale, filamentary structures. Indeed, this effect is most pronounced in $EE$, as expected from this interpretation. We defer a more detailed investigation of the small scale structure to future work.

One of the most surprising findings about Galactic polarized dust emission is that it has roughly twice as much power in $E$-mode polarization as $B$-mode \citep{Planck_Int_XXX}. The prevailing physical explanation is that the preferential elongation of density structures along the local magnetic field direction is responsible for this excess $E$-mode polarization, as well as for the observed positive $TE$ correlation \citep{Clark:2015,PlanckXXXVIII, Huffenberger:2019}. Indeed, \citet{Clark:2015} found that the $E/B$ asymmetry persisted for polarization templates built solely from maps of \hi orientation, without including polarized intensity information.

The $E/B$ asymmetry is readily apparent in Figure~\ref{fig:cross_spectra}, with $BB/EE \simeq 0.6$ on large angular scales ($\ell \lesssim 120$). We also compute the $BB/EE$ ratio derived from power-law fits to the \textit{Planck} 353\,GHz half-mission split cross-power spectra over the same mask and $\ell$ range, and likewise find $BB/EE \sim 0.6$ (see bottom panel of Figure~\ref{fig:cross_spectra}). No free parameters were adjusted to achieve this correspondence, strongly suggesting that the observed asymmetry indeed arises from filamentary density structures that are oriented along magnetic field lines. The ratio found here is comparable to, but slightly higher than, the \textit{Planck} value observed over large sky areas, $BB/EE = 0.53\pm0.01$ \citep{Planck2018XI}. The sky-variable $BB/EE$ ratio may be a useful probe of the physics of the dusty ISM, perhaps related to local properties of turbulence \citep[e.g.][]{Caldwell:2017}.

We quantify the scale-dependent correlation of the \HI-based maps with the {\it Planck} 353\,GHz maps in Figure~\ref{fig:r_planck}. As described in Section~\ref{subsec:data_dust}, we analyze the 353\,GHz {\it Planck} half-mission maps that have been CMB subtracted and smoothed to a resolution of $16.2'$. We plot the correlation ratios $r$ defined as

\begin{equation}\label{eq:corrratio}
    r^{XX}_{\hi \times 353} \equiv \frac{\mathcal{D}_\ell^{X_\hi X_{353}}}{\sqrt{\mathcal{D}_\ell^{X_\hi X_\hi} \times \mathcal{D}_\ell^{X_{353} X_{353}}}}
    ~~~,
\end{equation}
where the subscripts \hi and 353 denote the \hi and {\it Planck} 353\,GHz maps, respectively, and $X$ denotes either $E$ or $B$. The correlation between the maps is largest at large scales, with $r > 0.3$ for all $\ell < 100$ and reaching values as high as 0.75 and 0.60 for $EE$ and $BB$, respectively. For both cross spectra, $r$ declines roughly monotonically toward small scales where, as was noted in Figure~\ref{fig:cross_spectra}, the \hi maps appear to have more power than observed in the dust maps.

\section{Comparison to other tracers of interstellar magnetism}\label{sec:otherdatacomp}

The ISM is multi-phase, and a complete understanding of the interstellar magnetic field requires tracers beyond neutral hydrogen. Here we make a few connections to investigations of magnetic field structure using other tracers: radio polarimetric emission and optical starlight polarization. This is far from an exhaustive investigation, and we emphasize that there is much to be learned by comparing our three-dimensional Stokes maps to these and other data sets.

\subsection{LOFAR polarimetric filaments}

Observations of diffuse synchrotron emission trace the Faraday-rotated magneto-ionic medium. A new generation of radio polarimetric observations is opening a window into magnetic fields in the ionized interstellar medium, with facilities like the Low Frequency Array \citep[LOFAR;][]{vanHaarlem:2013}, the Murchison Wide-field Array \citep[MWA;][]{Tingay:2013}, and the multi-instrument project to map the full sky, the Global Magneto-Ionic Medium Survey \citep[GMIMS;][]{Wolleben:2019}. 

\begin{figure}
\centering
\includegraphics[width=
\columnwidth]{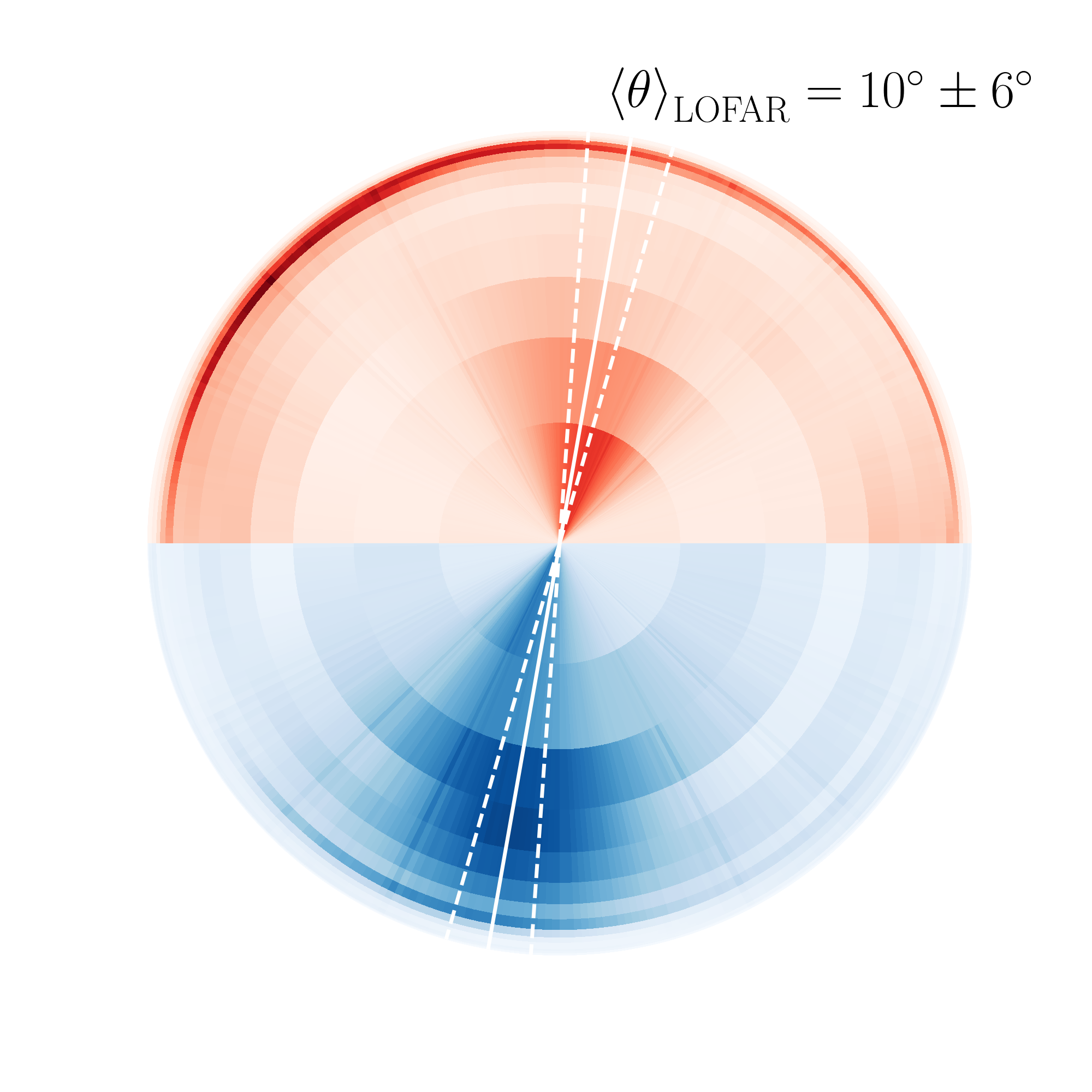}
\caption{Distribution of $I(v)R(v, \theta)$ in a circular region of $5^\circ$ diameter centered on 3C 196 ($l=171^\circ$, $b=33^\circ$). The most prominent orientation of the LOFAR depolarization canals is $\left<\theta\right>_\mathrm{LOFAR}=10^\circ \pm 6^\circ$ with respect to the Galactic Plane, according to the analyis by \citet{Jelic:2018}. The solid white line indicates $\left<\theta\right>=10^\circ$, and the dotted white lines denote the associated uncertainty.  }\label{fig:LOFARregion}
\end{figure}

LOFAR observations of a field centered on the quasar 3C~196 revealed strikingly linear, coherent structures in Faraday depth, including one prominent filament several degrees in length \citep{Jelic:2015}. These magneto-ionic structures are aligned with \hi filaments in the same field \citep{KalberlaKerp:2016, Jelic:2018}, as well as the \textit{Planck} 353\,GHz magnetic field orientation \citep{Zaroubi:2015}. The LOFAR observations are sensitive to magnetic structure in the warm ionized medium, while the dust and \hi are components of the neutral phase. Why does a morphological similarity persist across such different phases of the ISM? \citet{Jelic:2018} propose that the various ISM phases are confined by a well-ordered magnetic field that lies predominantly along the plane of the sky. The authors also observe alignment of \hi filaments across many velocity channels, indicating coherence of the magnetic field orientation along the line of sight \citep{Clark:2018}.

Here we examine the \citet{Jelic:2018} picture in the context of our three-dimensional Stokes maps. Figure~\ref{fig:LOFARregion} shows the sum of $I(v)R(v,\theta)$ over a circular region of diameter $5^\circ$, centered on 3C~196 at $(l, b) = (171^\circ, b=33^\circ)$. Our maps show a remarkably coherent orientation of linear \hi structures across a range of $v_{los}$, with a dominant \hi orientation ($\left<\theta_{\HI}\right> \sim 4^\circ$ over the full velocity range) that is consistent with the orientation of the Faraday depolarization canals measured by \citet{Jelic:2018} ($\left<\theta\right>_\mathrm{LOFAR}=10^\circ \pm 6^\circ$).\footnote{For consistency with \citet{Jelic:2018}, we report these orientation angles with respect to the Galactic Plane.} If the magnetic field is coherent and mostly in the plane of the sky, we expect our maps to have a high $p_{\HI}$ and a low $\mathcal{S}_{\HI}$ in this region. Indeed, averaging our \HI-based Stokes parameters over this circular region, we find $\left<p_{\HI}\right> = 11.5 \%$, higher than the full-sky average $\left<p_{\HI}\right> = 9.2 \%$. $\mathcal{S}_{\HI}$ is likewise low ($5.3^\circ$) over this region. \citet{Jelic:2018} found that EBHIS \hi filaments were oriented in the same direction in this field over $-11.5 < v < +3$\,\kmsa. Not surprisingly, Figure~\ref{fig:HIbinning} shows that the \HI4PI data (derived from EBHIS data in this region) are coherent in orientation-space over this velocity range, with a mean orientation $\left<\theta_{\HI}\right> \sim 12^\circ$. If we restrict our sums in Equations \ref{eq:sumI} - \ref{eq:sumU} to approximately this same velocity range, we find for this region $\left<p_{\HI}\right> = 46.7 \%$ (see discussion of $p_{\HI}$ in Section~\ref{sec:polfrac}). The mean \textit{Planck} polarization fraction in this region is $\left<p_{353}\right> = 9.6\%$.

Faraday depth, like \hi velocity, is not a direct probe of distance. However, this work and \citet{Jelic:2018} suggest that velocity- and Faraday depth-resolved probes of the magnetic field structure can be used in tandem to constrain the relative distances to magnetized structures. In conjunction with observations that more directly map to distance, these data can be used to map the magnetic field in three spatial dimensions, as well as to better understand the relationship between different phases of the magnetic ISM. Tomographic analyses combining probes of the neutral ISM with LOFAR observations \citep{vanEck:2017} and GMIMS observations \citep{Thomson:2019} have already been successful in selected regions of sky. We discuss further possibilities for three-dimensional magnetic tomography below.

\subsection{Starlight polarization}

\begin{figure*}
\centering
\includegraphics[width=
\textwidth]{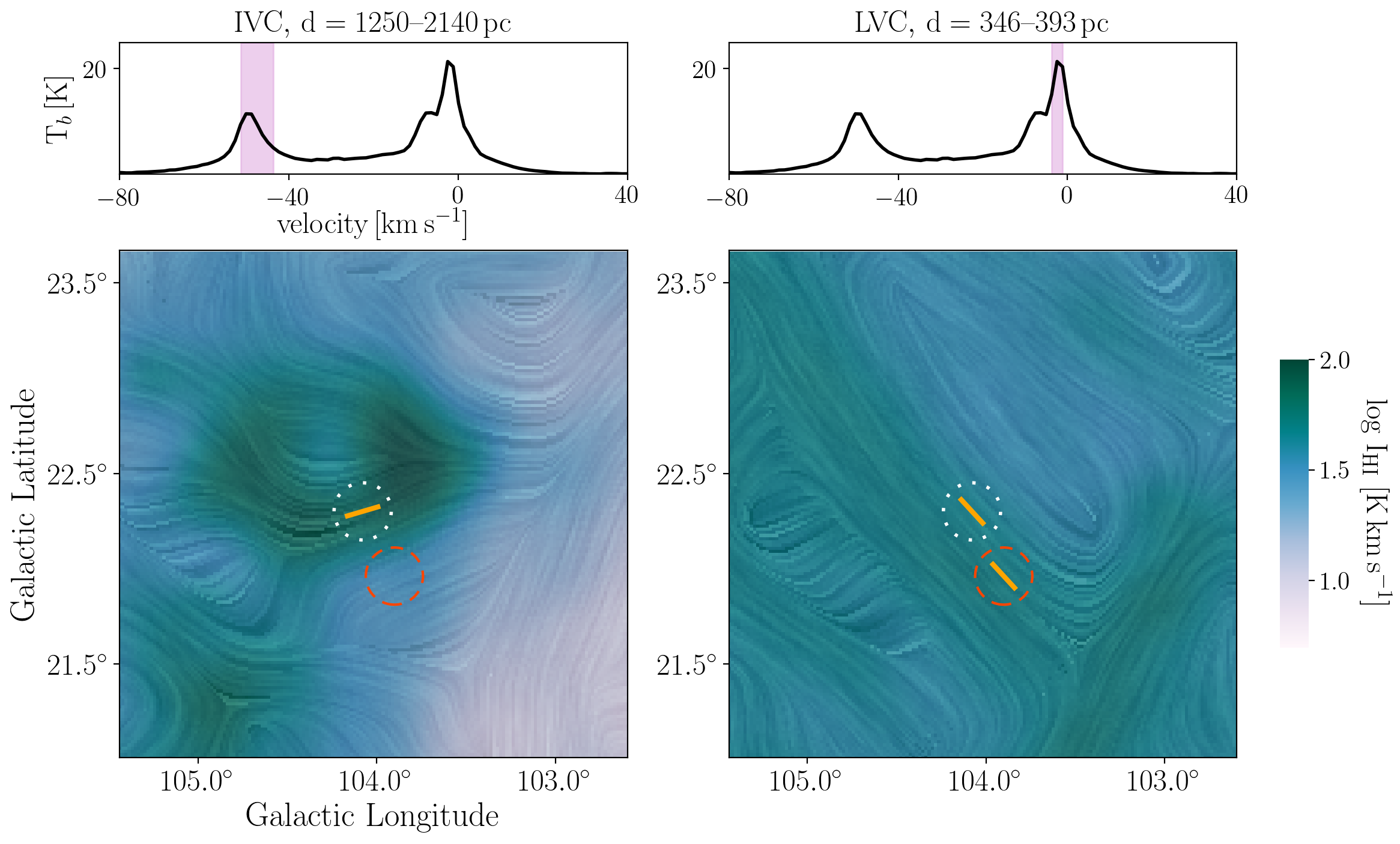}
\caption{A comparison between our velocity-resolved maps and the distance-decomposed magnetic field orientation estimate from \citet{Panopoulou:2019Tomo}. In their tomographic analysis, \citet{Panopoulou:2019Tomo} used starlight polarization measurements in the ``two-cloud" region indicated by the white dotted line, a line of sight that contains \hi emission from two prominent features, an intermediate velocity cloud (IVC) and a local velocity cloud (LVC). Their analysis also utilized a ``control" region with mostly local emission, indicated by the dashed red line. The mean \HI4PI spectrum over the two-cloud region is shown in the top two panels, with the velocity range associated with each cloud highlighted. The two maps show $I_{\HI}$ integrated over the indicated velocity regions, in a logarithmic colorbar stretch. The texture overlaid on the maps is our map of $\theta_{\HI}$ at $30'$, computed from $Q_{\HI}$ and $U_{\HI}$ integrated over the velocity regions highlighted in the top panels and visualized using line integral convolution \citep{Cabral:1993}. Orange line segments show the \citet{Panopoulou:2019Tomo} measurements of the local magnetic field orientation: those authors find that the IVC is located at a distance of $1250-2140\,\mathrm{pc}$ and has a mean magnetic field orientation $\left<\theta_\star\right> = 106^\circ \pm 8^\circ$, and the LVC is at a distance of $346-393\,\mathrm{pc}$ with $\left<\theta_\star\right> = 42.6^\circ \pm 1^\circ$. Averaged over the two-cloud region, we find $\left<\theta_{\HI}\right> = 111.6^\circ$ for the IVC and $\left<\theta_{\HI}\right> = 42.6^\circ$ for the LVC, in excellent agreement with the \citet{Panopoulou:2019Tomo} values.
}\label{fig:Ginaregion}
\end{figure*}

Starlight polarization probes the plane-of-sky magnetic field orientation in the dusty medium between observer and star \citep{Davis:1951}. As an integral probe, starlight polarization can be used to estimate the incremental polarization as a function of distance, given a sufficient density of polarimetric measurements and accurate stellar distances \citep[e.g.,][]{Lloyd:1973}. Recently, \citet{Panopoulou:2019Tomo} applied this principle to optical starlight polarimetry obtained with RoboPol \citep{Ramaprakash:2019}, for stars with distances estimated from \textit{Gaia} parallax measurements \citep{GaiaDR2:2018, BailerJones:2018}. 

\citet{Panopoulou:2019Tomo} selected a circular region of sky of radius $0.16^\circ$, centered on $(l, b) = (104.08^\circ, 22.31^\circ)$. Dubbed the ``two-cloud" region, this sightline passes through two \hi emission structures that are distinct in velocity space: an intermediate velocity cloud (IVC) and a local velocity cloud (LVC; Figure \ref{fig:Ginaregion}). The polarimetric measurements toward stars in this region are compared to those in a control region of the same angular size centered on $(l, b) = (103.90^\circ, 21.97^\circ)$ that contains less \hi emission from intermediate velocities. The mean starlight polarization angles for stars that lie in each of these two regions agree within the error bars. However, \citet{Panopoulou:2019Tomo} use stellar distances and the on-off setup of their selected regions to carefully disentangle the polarimetric properties associated with the near and far clouds. The authors find that the IVC is farther away, at a distance of $1250-2140\,\mathrm{pc}$, and has a mean magnetic field orientation\footnote{For consistency with the rest of this work, we report these measurements in the IAU Galactic polarization convention, converted from their reported Equatorial frame coordinates following \citet{Appenzeller:1968}. See also \citet{Panopoulou2016Erratum}.} $\left<\theta_\star\right> = 106^\circ \pm 8^\circ$. For the LVC they find a distance of $346-393\,\mathrm{pc}$ with $\left<\theta_\star\right> = 42.6^\circ \pm 1^\circ$.

To compare our three-dimensional Stokes parameter maps with the \citet{Panopoulou:2019Tomo} polarimetric analysis, we select the velocity range in our maps closest to the ranges associated with the IVC and LVC: these are $-51.4 < v < -38.5$\,\kms and  $-3.8 < v < -1.2$\,\kmsa, respectively. We compute $Q_{\HI}$ and $U_{\HI}$ from Equations~\ref{eq:sumQ} and \ref{eq:sumU} over these velocity ranges, and visualize the resulting magnetic field orientation $\theta_{\HI}$ in the two regions in Figure~\ref{fig:Ginaregion}. We compute the mean magnetic field orientation in the two-cloud region defined above, finding $\left<\theta_{\HI}\right> = 111.6^\circ$ for the IVC and $\left<\theta_{\HI}\right> = 42.6^\circ$ for the LVC in our maps at $30'$. These values agree well with the polarimetry-derived measurements for the two clouds. We find a higher $\left<p_{\HI}\right>$ for the Stokes maps integrated over the LVC velocity range than for the IVC velocity range, just as \citet{Panopoulou:2019Tomo} derive a higher fractional linear polarization associated with the LVC than with the IVC.

It is worth emphasizing the remarkable correspondence between our three-dimensional Stokes maps and the \citet{Panopoulou:2019Tomo} measurements. Our maps probe the local magnetic field as a function of velocity. Starlight polarization measures properties of the integrated magnetic field, which can be used in conjunction with stellar distance measurements to probe the magnetic field as a function of distance. The implications for higher-dimensional tomography of the magnetic ISM are discussed further in Section~\ref{sec:discussion}.

\section{Model Variations and extensions}\label{sec:variations}

\subsection{Gradient-based determination of \hi orientation}\label{sec:gradient}

Mapping magnetic coherence requires some method of quantifying the orientation of \hi structure. In this work we use the RHT, but our framework is not algorithm-specific. The RHT is not the only conceivable mapping from image space to orientation space, and there are a range of reasonable parameter choices for the RHT that could be adopted. Such differences would change the numerical values of $Q(v)$ and $U(v)$, although these are circumscribed by $I(v)$, which is directly measured from the sky, and by the fact that the derived orientation in a given \hi region (the ratio of $Q$ and $U$) tends to be fairly insensitive to RHT parameter choices \citep[][]{Clark:2014}. We consider the maps provided here to be an excellent estimate of the three-dimensional Stokes parameters derived from \hi emission, but not a unique solution. Another way to measure the orientation of image structures is simply to take a spatial gradient, and to define the direction of maximum gradient as the local feature orientation. The spatial gradient is widely used in machine vision for edge detection, feature matching, and more complex applications, and has been used in astrophysical contexts to compare ISM structure to the magnetic field orientation \citep{Soler:2013}.

Taking a spatial gradient is a computationally simpler procedure than the RHT, but it collapses any information about the distribution of \hi emission in orientation space. If this distribution is simply summed along the $\theta$ axis as in Equations~\ref{eq:QHIv} and \ref{eq:UHIv}, then the final orientation angle will be similar as long as the local gradient orientation is similar to the mean RHT angle. A major caveat to this is that the RHT measures linearity as a function of orientation, and pixels will have zero $R(v, \theta)$ if there is insufficient linear intensity. This is a desirable property for building three-dimensional Stokes maps if the prominently linear \hi structures are the most predictive of the polarized intensity. In contrast, the spatial gradient measures the local direction of maximum change in the \hi intensity without regard for the feature that produced it. A circular blob of \hi intensity could contribute nothing to $R(v, \theta)$, but have strong spatial gradients oriented in all directions around its edge. Without further processing, the spatial gradient is thus very sensitive to small-scale noise and other data features that are not predictive of magnetic field properties. This property is evident in the spatial gradient of \HI4PI channel maps. Because the EBHIS and GASS surveys have different angular resolutions, one hemisphere was convolved with a larger beam to create the \HI4PI dataset at uniform resolution. The noise properties thus differ across the map, and this manifests in starkly different properties of the spatial gradient between the two hemispheres. Furthermore, the mapping to velocity-orientation space enabled by the RHT is a natural description of magnetic coherence that allows further exploration of dust properties (see Section~\ref{sec:discussion}).

We create an alternative set of three-dimensional Stokes maps using the spatial gradient of the \hi emission $I_{\HI}(l, b, v)$. If the \hi emission in a given velocity channel is defined on the sphere as $I_{\HI}(\theta, \phi)$, we take 

\beq
\nabla I_{\ion{H}{i}} = 
\begin{bmatrix} 
I_{\theta} \cr 
I_\phi 
\end{bmatrix} 
= 
\begin{bmatrix}
\partial_\theta I_{\HI} \cr
\frac{1}{\sin\theta} \partial_\phi I_{\HI}
\end{bmatrix}
\eeq
and define the local feature orientation as orthogonal to the direction of the steepest intensity gradient,

\beq 
\theta_\mathrm{grad} = \arctan \left( \frac{I_\theta}{I_\phi} \right).
\eeq
In this approach as with the RHT, we use an orientation convention that corresponds to the polarization convention used, such that the sense of the \hi orientation is a direct proxy for the projected magnetic field orientation, which is orthogonal to the polarization angle of dust emission. We then compute 

\begin{align}
    Q_\ion{H}{i}^\mathrm{grad}\left(v\right) &= I\left(v\right) \cos\left(2\theta_\mathrm{grad}\right)\label{eq:QHIv_gradient}\\ 
    U_\ion{H}{i}^\mathrm{grad}\left(v\right) &= I\left(v\right) \sin\left(2\theta_\mathrm{grad}\right)\label{eq:UHIv_gradient},
\end{align}
and Equations~\ref{eq:sumI} - \ref{eq:sumU} hold, as in the RHT-based construction of the \hi Stokes parameters. We then have $Q_{\HI}^\mathrm{grad}$, $U_{\HI}^\mathrm{grad}$, and the derived quantities $p_{\HI}^\mathrm{grad}$ and $\theta_{\HI}^\mathrm{grad}$, which can be compared to both the \textit{Planck} 353\,GHz measurements and the RHT-based $Q_{\HI}$ and $U_{\HI}$ quantities considered in the rest of this work. 

We find that the gradient-based Stokes parameters are similar to the RHT-derived values, but the RHT-derived maps are better correlated with the \textit{Planck} measurements. For instance, a simple linear regression of $Q_{\HI}$ vs. $Q_{\HI}^\mathrm{grad}$ at $80'$ yields a correlation coefficient $r=0.91$. This value is $r=0.87$ for $Q_{\HI}$ vs. $Q_{353}$, and $0.82$ for $Q_{\HI}^\mathrm{grad}$ vs. $Q_{353}$. We find similar results when comparing derived quantities across the three sets of maps: $\xi=0.86$ between $\theta_{\HI}$ and $\theta_{\HI}^\mathrm{grad}$, $\xi=0.71$ between $\theta_{\HI}$ and $\theta_{353}$, and $\xi=0.67$ between $\theta_{\HI}^\mathrm{grad}$ and $\theta_{353}$. Likewise, $p_{353}$ is better correlated with $p_{\HI}$ than with $p_{\HI}^\mathrm{grad}$. The gradient-based maps have a lower correlation ratio (Equation~\ref{eq:corrratio}) with the 353\,GHz maps than the RHT-based maps do ($r_\ell^{EE}$ with the gradient is $\lesssim r_\ell^{EE}$ with the RHT for all $\ell$).

It is further instructive to compare the small-scale structure of the gradient-based maps to the maps constructed using the RHT. The righthand panel of Figure \ref{fig:smallareaPHI} shows $P_{\HI}$ constructed using Equations~\ref{eq:QHIv_gradient} and \ref{eq:UHIv_gradient}. Because the gradient is computed over small scales, the polarized intensity is not concentrated into structures as coherently linear as those in the RHT-based maps. Prominent \hi filaments become bright $P_{\HI}$ filaments in the RHT-based maps, but it is often the edges of those filaments that contain strong gradients and thus show up as coherent structures in the $P_{\HI}^\mathrm{grad}$ map.

\subsection{The relative orientation between \hi structures and the magnetic field}\label{sec:thetadiff}

In the diffuse ISM, the orientation of \hi structures traces the plane-of-sky magnetic field with high fidelity. A fundamental assumption of the maps presented here is that the orientation of \hi channel map structures is predictive of the magnetic field orientation over the whole sky. In other tracers, notably dust \citep{PlanckXXXII} and molecular line emission \citep[][]{Goldsmith:2008}, there is evidence for a loss of alignment between structures and the magnetic field toward high-column sightlines. We find no evidence for a global decrease in the alignment between $\theta_{353}$ and $\theta_{\HI}$ at $80'$ as a function of \nHI in the \HI4PI data. Similar results were obtained for the correlation between $\theta_{353}$ and $\theta_{\HI}^\mathrm{grad}$. We therefore find no motivation to introduce an \nhi-dependent expectation for the relative orientation between the \hi structures and the local magnetic field. Any loss of relative alignment between the \hi structures and the magnetic field orientation at high \nHI is likely only detectable on scales smaller than the $80'$ \textit{Planck} resolution that we have used in this analysis. In contrast, we note that the dependence of $\xi$ on $\mathcal{S}_{\HI}$ is a significant effect (see Figure~\ref{fig:deltatheta_S}).

\subsection{\hi phase decomposition}
Galactic \hi exists in multiple phases, and any given sightline may contain emission from the CNM, the warm neutral medium (WNM), and gas in a thermally unstable phase \citep[e.g.,][]{Cox:2005}. As discussed above, the \hi structures that are most predictive of the magnetic field orientation are predominantly CNM. In fact, the small-scale structure in \hi channel maps in general is preferentially cold gas \citep{Clark:2019}. By construction, our maps assign polarized intensity to these structures in a given velocity bin, but only total intensity to the more diffuse, inter-structure gas. In reality, all \hi intensity is probably correlated with polarized dust emission at some level, and it may be that our maps currently over-weight the contribution of the CNM to the polarized emission from the neutral medium. Because the WNM-associated \hi emission does not strongly contribute to the \hi orientation measurements, the empirical correlation between our \HI-based maps and the 353\,GHz dust polarization constrains the magnetic coherence between the WNM and the CNM.

Phase-decomposed maps of \hi emission can be used to explicitly assign different emission properties to different \hi phases or velocity ranges, a strategy employed in the dust models of \citet{Ghosh:2017} and \citet{Adak:2019}, and in the foreground cleaning for CIB maps of \citet{Lenz:2019}. The relationship between the phase structure of \hi and the ambient magnetic field is worth further exploration, especially as sophisticated Gaussian fitting algorithms for phase-decomposing hyperspectral data are being developed \citep[e.g.][]{Marchal:2019, Riener:2019}.

\subsection{Connecting \hi and $\gamma$}\label{sec:HIgamma}

We establish in this work that regions of high polarization angle dispersion ($\mathcal{S}$) are regions where $\theta_{\HI}$ is less well correlated with $\theta_{353}$ (Figure~\ref{fig:deltatheta_S}). We find a similar relationship between $\mathcal{S}_{\HI}$ and $p_{\HI}$ to the one found between these quantities in the \textit{Planck} 353\,GHz maps (Figure~\ref{fig:pvsS2dhist}). These observations are qualitatively consistent with a picture in which $\mathcal{S}$ is sensitive to the angle $\gamma$ between the mean magnetic field and the plane of the sky \citep{FalcetaGoncalves+etal_2008,Planck_Int_XX, King:2018}. When the magnetic field is along the line of sight, both small perturbations to its 3D orientation, e.g., from interstellar turbulence, and instrumental noise in the $Q$ and $U$ maps induce large changes to its projected 2D orientation, thereby leading to large values of $\mathcal{S}$. In contrast, when the magnetic field lies mostly in the plane of the sky, perturbations to its 3D orientation have little effect on the projected 2D orientation, leading to small values of $\mathcal{S}$. \citet{Hensley:2019} recently provided empirical evidence for this picture by finding that the dust emission per $N_\HI$ is positively correlated with $\mathcal{S}$, as expected from orientation effects.

The dust polarization fraction depends in large part on the value of $\gamma$. When the magnetic field is along the line of sight, grain rotation about the field results in no net polarization. When the magnetic field lies in the plane of the sky, the polarization fraction is maximal. In the modeling framework presented here, we do not explicitly account for the effects of $\gamma$ on the polarized emission. Yet we find that $p_\HI$ is remarkably well correlated with $p_{353}$ (see Figure~\ref{fig:pvsp2dhist}). To understand this, we note that $Q_\HI$ and $U_\HI$ are constructed by integrating $R(v, \theta)$ over orientation. In regions where the dispersion in polarization angles is high, this results in strong depolarization. These regions of high $\mathcal{S}$ are also the ones where, on average, the magnetic field is expected to be oriented more along the line of sight, and the dust polarization fraction is consequently low. On the other hand, regions with low $\mathcal{S}$ are not significantly depolarized when integrating $R(v, \theta)$ over $\theta$, resulting in high polarization fractions. Thus the effect of $\gamma$ is implicitly encoded in $p_\HI$. Explicitly modeling $\gamma$ from the \HI, perhaps by introducing a factor inversely proportional to $\mathcal{S}_{\HI}$ to Equations~\ref{eq:QHIv} and \ref{eq:UHIv}, could improve the correlation between these maps and the 353\,GHz polarized dust maps. 

Relatedly, this work suggests that the velocity structure of \hi can be used to decompose the two geometric effects that contribute to depolarization: plane-of-sky depolarization, arising from dispersion of polarization angles within the beam, and line-of-sight depolarization, from multiply oriented dust polarization angles along the line of sight. These effects can in principle be isolated and quantified with our 3D Stokes maps.

\subsection{Other considerations}
The present maps are constructed using a fixed RHT window size, meaning that the minimum angular length of \hi structures that contribute to $\theta_{\HI}$ does not vary as a function of velocity. Because a fixed angular size corresponds to different physical sizes for structures at different distances, maps integrated over a large velocity range likely incorporate structures of different physical scales: specifically, more distant structures would need to be physically longer in order to be detected as linear features. This may not be optimal for some use cases, and could in principle be adjusted, given distance information to the gas. We note, however, that any scale-mixing from this effect is likely unimportant at high Galactic latitudes, where the dust is concentrated no farther than a few hundred parsecs away \citep{Capitanio:2017}.

We do not find any significant loss of correlation between our maps and the \textit{Planck} 353\,GHz dust $Q$ and $U$ maps as a function of Galactic latitude. This is perhaps surprising, as the previous work that motivates our magnetically coherent mapping (Section~\ref{sec:obsbasis}) was mostly focused on high Galactic latitudes. Evidently \hi orientation remains a reliable proxy for magnetic field orientation along sightlines closer to the Galactic Plane, although it would be interesting to investigate whether the structures that dominate this measurement are substantially different physically from the ``\hi fibers" \citep{Clark:2014} that pattern the diffuse sky.

\section{Discussion}\label{sec:discussion}

\subsection{Relevance for foreground mitigation}

In addition to the study of the ISM, our model has significant relevance to cosmology. One of the major goals of modern observational cosmology is the search for primordial $B$-mode polarization in the CMB, a signature of inflation-era gravitational waves \citep{Kamionkowski:1997, Seljak:1997}. A paramount obstacle for a successful detection is the polarized foreground signal from Galactic dust emission, which at CMB frequencies is several orders of magnitude brighter in polarization than the sought-after primordial signal \citep{Dunkley:2009, Flauger:2014, BICEPPlanckJoint2015, Planck2018XI}. 

Realistic models of Galactic polarized dust emission are therefore a crucial ingredient for experiment design and forecasting, component separation, and validation of any claimed detection. Polarized CMB foregrounds arise from the three-dimensional ISM, which is shaped by nonlinear MHD processes that are fundamentally non-Gaussian, particularly on smaller scales \citep[e.g.,][]{Burkhart:2009, Allys:2019}. While data-driven sky models like the Planck Sky Model \citep{Delabrouille:2013} or the Python Sky Model \citep{Thorne:2017} are able to incorporate this non-Gaussianity on scales that have already been well measured, they typically resort to simple Gaussian parameterizations at small scales.

A number of approaches to provide this small-scale information are being actively pursued. MHD simulations have successfully reproduced many of the observed properties of the polarized dust emission \citep{Kritsuk:2018, Kim:2019}, and, if pushed to sufficiently high resolution, can generate small-scale, non-Gaussian spatial fluctuations in a physically motivated way. However, MHD simulations are limited in being able to reproduce only the statistical properties of the Galaxy, not the specific morphology of the Galaxy itself. High-resolution ancillary data like Galactic \hi also naturally contain physical, non-Gaussian spatial structure while also corresponding directly to the particular structure of our Galaxy. Past efforts have largely focused on two-dimensional data with a parametric, phenomenological perscription for the line-of-sight structure \citep[e.g.,][]{PlanckXLIV, Vansyngel:2017, Ghosh:2017, Adak:2019}. While simple, these models have been effective in reproducing many of the spatial statistics of the polarized dust emission, such as the $E/B$ asymmetry and polarization cross spectra. Rather than employ a phenomenological prescription for line-of-sight structure, our maps incorporate the three-dimensional distribution of magnetic coherence entirely from data.

The maps presented in this work also provide a data-driven framework for modeling spatial variability in the frequency-dependent dust emission. The superposition of multiple emitting regions along the line of sight, each with their own dust properties (such as temperature) and magnetic field orientation, hinders the ability of a map of dust emission at one frequency to be extrapolated to another frequency, i.e., ``frequency decorrelation'' \citep{TassisPavlidou:2015}. While currently not detected in the {\it Planck} dust polarization data \citep{Planck2018XI}, even modest levels of decorrelation can impact B-mode science \citep{Poh:2017, HensleyBull:2018}.

Line-of-sight variability of the dust emission has been modeled both with simple parametric prescriptions that reproduce the 2D observations as well as directly incorporating 3D data sets such as dust extinction maps \citep{Poh:2017, MartinezSolaeche:2018}. Our maps can improve considerably on these methods by incorporating not only line-of-sight information of the density distribution, but also the magnetic field orientation. Mapping different dust emission models onto the three-dimensional maps derived in this work will provide new data-driven models and predictions for the expected levels of frequency decorrelation.

Finally, we note that even as-is, the \HI-based Stokes maps are an ideal template for assessing the residual contamination in a foreground-cleaned CMB observation \citep[e.g.,][]{Thorne:2019}. Any cross-correlation signal between such maps and our model is certain to arise from the Galactic ISM. Because we restrict the \hi data to velocities $|v| < 90$\,km\,s$^{-1}$, our \HI-based maps are free from CIB contamination \citep[][]{Chiang:2019}. Further, as an entirely independent data set, the \HI-derived maps will not contain correlated systematics with any CMB observations.

\subsection{Relevance for global magnetic field modeling}

The global structure of the Galactic magnetic field (GMF) is an outstanding problem in astrophysics, with wide-ranging consequences for understanding cosmic ray propagation, dynamo theory, star formation, and galactic evolution \citep{Haverkorn:2015, Jaffe:2019}. At present, the state-of-the-art data-driven models for the Galactic magnetic field are parametric fits to a few observables, typically some subset of extragalactic and pulsar rotation measures (RMs), synchrotron emission, and polarized dust emission \citep[e.g.][]{JanssonFarrar:2012, Jaffe:2013, Terral:2017}. Each model relies on a set of assumptions and contains large uncertainties, such that there is a wide variation in the GMF structure obtained, even between models based on the same observational data. New observational constraints and more robust statistical frameworks are two avenues for improvement \citep{Boulanger:2018, Haverkorn:2019}. The \HI-derived Stokes maps constitute a new observational constraint, and here we detail a few specific ideas for their use in GMF modeling.

The influence of nearby ($\lesssim\mathrm{few}\,\mathrm{hundred}\,\mathrm{pc}$) ISM structure on observational tracers presents a major challenge for GMF models. Our proximity to these local features means that they may be observed on large angular scales, but may be unimportant to the Galactic-scale field structure. For instance, the Sun sits in the Local Bubble (also called the Local cavity), an underdense region of the nearby ISM thought to have been carved out by a series of supernovae \citep{Cox:1987, Lallement:2003}. The expansion of the Bubble likely distorted the local GMF, measurably impacting the observed dust emission, and yet the Local Bubble is not included in GMF models \citep{Alves:2018}. The 3D Stokes maps presented here can be used to disentangle the influence of these local features from the rest of the emission in observations like the \textit{Planck} 353\,GHz map. Our maps may be particularly interesting to compare with detailed 3D maps of the local ISM \citep{Lallement:2014, Capitanio:2017}, as well as with the \citet{Alves:2018} model for the magnetic structure of the Local Bubble.

Faraday rotation is an important and widely used probe of the GMF \citep[e.g.,][]{Hutschenreuter:2019}. RM is an integral quantity, proportional to the product of the thermal electron density ($n_e$) and the line-of-sight component of the magnetic field integrated between the observer and the source. The largest uncertainty in estimating the magnetic field strength from measurements of RM is thus the distribution of $n_e$ along the line of sight \citep{Beck:2003}. A better understanding of the magnetic field and its relationship to gas density between phases of the ISM is crucial to making progress \citep{Boulanger:2018}. If our maps primarily trace the CNM, they may be useful for mapping the three-dimensional phase distribution of the ISM, or for identifying sharp changes along the line of sight in either $n_e$ or the magnetic field, in the vein of \citet{vanEck:2017}.

\subsection{Toward four-dimensional magnetic tomography}

Ultimately, we see this work as an important step toward the goal of mapping the three-dimensional structure of the Galactic magnetic field in three spatial dimensions. Our 3D Stokes maps constitute a novel measurement: a local probe of the magnetic field orientation as a function of velocity. Other measurements of the magnetic field orientation in the neutral ISM are not local\footnote{Zeeman splitting is a local probe of the magnetic field, but is not currently practical for mapping the diffuse ISM.}, but integrated: polarized dust emission traces the field integrated along the entire line of sight, and starlight polarization traces the field between the observer and the star.

The next step is to map these data to distance. This calls for the synthesis of maps like those presented in this work with other data sets. Stellar reddening and other measurements enable maps of the dust distribution in three spatial dimensions \citep[e.g.,][]{Marshall:2006, Green:2019}. Large-scale starlight polarization surveys like PASIPHAE at high Galactic latitudes \citep{PASIPHAE:2018} will soon provide optical polarimetry toward stars with distances measured by \textit{Gaia} \citep{Gaia:2016}, improving the density of starlight polarization measurements at high Galactic latitude by orders of magnitude over existing catalogs \citep[e.g.,][]{Heiles:2000, Berdyugin:2014}. These data sets can be combined with our velocity-resolved maps to produce maps in four dimensions -- three spatial dimensions and radial velocity. Some of these data have already been used to map the four-dimensional density distribution in the Galactic Plane \citep{Tchernyshyov:2017}.

\section{Conclusions}\label{sec:conclusions}
We summarize the principal conclusions of this work below.

\begin{itemize}
    \item We define a paradigm for mapping magnetic coherence from hyperspectral observations of Galactic neutral hydrogen. The principle of magnetic coherence is motivated by three empirical correlations between \hi and dust in the diffuse ISM: that \hi column traces the dust column \citep[e.g.,][]{Lenz:2017}, that \hi orientation traces the dust polarization angle \citep{Clark:2015}, and that the coherence of \hi orientation as a function of velocity traces the dust polarization fraction \citep{Clark:2018}.

    \item We construct three-dimensional maps of Stokes $I$, $Q$, and $U$, using only \hi data. These maps represent the interstellar magnetic field structure as a function of radial velocity. We construct full-sky maps using the $16.2'$ \HI4PI survey, and $4'$ partial-sky maps using GALFA-\hi data.
    
    \item Integrating our maps over the velocity dimension, we obtain two-dimensional maps of $I_{\HI}$, $Q_{\HI}$, $U_{\HI}$, which we quantitatively compare with \textit{Planck} measurements of the linearly polarized 353\,GHz Galactic emission. We likewise compute the plane-of-sky magnetic field orientation, polarization fraction, and polarization angle dispersion function for our velocity-integrated maps and compare them to the 353\,GHz data. We find striking agreement between our \HI-based maps and the \textit{Planck} measurements. This is particularly remarkable as our maps contain no free parameters, and we have not explicitly modeled any properties of dust emission.
    
    \item We find an anticorrelation between $p_{\HI}$ and $\mathcal{S}_{\HI}$, similar to the relationship measured for these quantities in the 353\,GHz data. The correlation between $\theta_{\HI}$ and $\theta_{353}$ depends strongly on $\mathcal{S}_{\HI}$: in regions of low polarization angle dispersion, the \HI-based magnetic field orientation is better aligned with the 353\,GHz magnetic field orientation. We discuss the sensitivity of $\mathcal{S}$ to the line-of-sight magnetic field orientation.
    
    \item We measure $TE$, $EE$, and $BB$ cross-power spectra of our \HI-based maps and discuss their relationship to the structure of the small-scale polarized emission in our maps. We find $BB/EE \sim 0.6$, in agreement with the \textit{Planck} measurement of this ratio in the region of sky considered. This strongly supports the interpretation that the $BB/EE$ asymmetry in the polarized dust sky is due to the preferential elongation of density structures along the local magnetic field. We also measure the correlation between the \HI-based and 353\,GHz $E$- and $B$-mode polarization maps, finding that these maps are highly correlated on large angular scales, with the correlation ratio declining toward high multipole.  
    
    \item Aside from the polarized dust emission, we compare our maps to two distinct tracers of magnetism: low-frequency radio polarimetric observations of the magneto-ionic medium, and optical starlight polarization measurements. Our maps support the physical explanation for the LOFAR Faraday depth filaments discussed in \citet{Jelic:2018}. Our \HI-based measurement of the \textit{local} magnetic field orientation along a line of sight dominated by two distinct dust components agrees extremely well with the \citet{Panopoulou:2019Tomo} tomographic decomposition of starlight polarization toward stars with distances measured by \textit{Gaia}. We discuss prospects for using our maps for magnetic tomography.

\end{itemize}

We will make the three-dimensional \HI-based Stokes parameter maps discussed in this work, as well as their velocity-integrated counterparts, publicly available upon publication. 

\software{astropy \citep{Astropy:2013, Astropy:2018}, Healpix \citep{Healpix:2005}, healpy \citep{Zonca:2019}, matplotlib \citep{Matplotlib:2007}, NaMaster \citep{Alonso:2019}, numpy \citep{Oliphant:2015:GN:2886196}}

\acknowledgments

We thank many colleagues for enlightening conversations over the course of this work, particularly Daniel Lenz, J. Colin Hill, Jo Dunkley, and Nicoletta Krachmalnicoff.

S.E.C. is supported by NASA through Hubble Fellowship grant \#HST-HF2-51389.001-A awarded by the Space Telescope Science Institute, which is operated by the Association of Universities for Research in Astronomy, Inc., for NASA, under contract NAS5-26555. 

This publication utilizes data from Galactic ALFA \hi (GALFA-\HI) survey data set obtained with the Arecibo L-band Feed Array (ALFA) on the Arecibo 305m telescope. The Arecibo Observatory is operated by SRI International under a cooperative agreement with the National Science Foundation (AST-1100968), and in alliance with Ana G. M\'endez-Universidad Metropolitana, and the Universities Space Research Association. The GALFA-\hi surveys have been funded by the NSF through grants to Columbia University, the University of Wisconsin, and the University of California.

This publication utilizes observations obtained with Planck (http://www.esa.int/Planck), an ESA science mission with instruments and contributions directly funded by ESA Member States, NASA, and Canada.

\bibliography{references}

\begin{thebibliography}{}
\expandafter\ifx\csname natexlab\endcsname\relax\def\natexlab#1{#1}\fi

\bibitem[{{Adak} {et~al.}(2019){Adak}, {Ghosh}, {Boulanger}, {Kalberla},
  {Haud}, {Martin}, {Bracco}, \& {Souradeep}}]{Adak:2019}
{Adak}, D., {Ghosh}, T., {Boulanger}, F., {et~al.} 2019, arXiv e-prints,
  arXiv:1906.07445

\bibitem[{{Allys} {et~al.}(2019){Allys}, {Levrier}, {Zhang}, {Colling},
  {Regaldo-Saint Blancard}, {Boulanger}, {Hennebelle}, \&
  {Mallat}}]{Allys:2019}
{Allys}, E., {Levrier}, F., {Zhang}, S., {et~al.} 2019, arXiv e-prints,
  arXiv:1905.01372

\bibitem[{{Alonso} {et~al.}(2019){Alonso}, {Sanchez}, {Slosar}, \& {LSST Dark
  Energy Science Collaboration}}]{Alonso:2019}
{Alonso}, D., {Sanchez}, J., {Slosar}, A., \& {LSST Dark Energy Science
  Collaboration}. 2019, \mnras, 484, 4127

\bibitem[{{Alves} {et~al.}(2018){Alves}, {Boulanger}, {Ferri{\`e}re}, \&
  {Montier}}]{Alves:2018}
{Alves}, M.~I.~R., {Boulanger}, F., {Ferri{\`e}re}, K., \& {Montier}, L. 2018,
  \aap, 611, L5

\bibitem[{{Appenzeller}(1968)}]{Appenzeller:1968}
{Appenzeller}, I. 1968, \apj, 151, 907

\bibitem[{{Asensio Ramos} {et~al.}(2017){Asensio Ramos}, {de la Cruz
  Rodr{\'{\i}}guez}, {Mart{\'{\i}}nez Gonz{\'a}lez}, \&
  {Socas-Navarro}}]{AsensioRamos:2017}
{Asensio Ramos}, A., {de la Cruz Rodr{\'{\i}}guez}, J., {Mart{\'{\i}}nez
  Gonz{\'a}lez}, M.~J., \& {Socas-Navarro}, H. 2017, \aap, 599, A133

\bibitem[{{Astropy Collaboration} {et~al.}(2013){Astropy Collaboration},
  {Robitaille}, {Tollerud}, {Greenfield}, {Droettboom}, {Bray}, {Aldcroft},
  {Davis}, {Ginsburg}, {Price-Whelan}, {Kerzendorf}, {Conley}, {Crighton},
  {Barbary}, {Muna}, {Ferguson}, {Grollier}, {Parikh}, {Nair}, {Unther},
  {Deil}, {Woillez}, {Conseil}, {Kramer}, {Turner}, {Singer}, {Fox}, {Weaver},
  {Zabalza}, {Edwards}, {Azalee Bostroem}, {Burke}, {Casey}, {Crawford},
  {Dencheva}, {Ely}, {Jenness}, {Labrie}, {Lim}, {Pierfederici}, {Pontzen},
  {Ptak}, {Refsdal}, {Servillat}, \& {Streicher}}]{Astropy:2013}
{Astropy Collaboration}, {Robitaille}, T.~P., {Tollerud}, E.~J., {et~al.} 2013,
  \aap, 558, A33

\bibitem[{{Astropy Collaboration} {et~al.}(2018){Astropy Collaboration},
  {Price-Whelan}, {Sip{\H o}cz}, {G{\"u}nther}, {Lim}, {Crawford}, {Conseil},
  {Shupe}, {Craig}, {Dencheva}, {Ginsburg}, {VanderPlas}, {Bradley},
  {P{\'e}rez-Su{\'a}rez}, {de Val-Borro}, {Aldcroft}, {Cruz}, {Robitaille},
  {Tollerud}, {Ardelean}, {Babej}, {Bach}, {Bachetti}, {Bakanov}, {Bamford},
  {Barentsen}, {Barmby}, {Baumbach}, {Berry}, {Biscani}, {Boquien}, {Bostroem},
  {Bouma}, {Brammer}, {Bray}, {Breytenbach}, {Buddelmeijer}, {Burke},
  {Calderone}, {Cano Rodr{\'{\i}}guez}, {Cara}, {Cardoso}, {Cheedella},
  {Copin}, {Corrales}, {Crichton}, {D'Avella}, {Deil}, {Depagne}, {Dietrich},
  {Donath}, {Droettboom}, {Earl}, {Erben}, {Fabbro}, {Ferreira}, {Finethy},
  {Fox}, {Garrison}, {Gibbons}, {Goldstein}, {Gommers}, {Greco}, {Greenfield},
  {Groener}, {Grollier}, {Hagen}, {Hirst}, {Homeier}, {Horton}, {Hosseinzadeh},
  {Hu}, {Hunkeler}, {Ivezi{\'c}}, {Jain}, {Jenness}, {Kanarek}, {Kendrew},
  {Kern}, {Kerzendorf}, {Khvalko}, {King}, {Kirkby}, {Kulkarni}, {Kumar},
  {Lee}, {Lenz}, {Littlefair}, {Ma}, {Macleod}, {Mastropietro}, {McCully},
  {Montagnac}, {Morris}, {Mueller}, {Mumford}, {Muna}, {Murphy}, {Nelson},
  {Nguyen}, {Ninan}, {N{\"o}the}, {Ogaz}, {Oh}, {Parejko}, {Parley}, {Pascual},
  {Patil}, {Patil}, {Plunkett}, {Prochaska}, {Rastogi}, {Reddy Janga},
  {Sabater}, {Sakurikar}, {Seifert}, {Sherbert}, {Sherwood-Taylor}, {Shih},
  {Sick}, {Silbiger}, {Singanamalla}, {Singer}, {Sladen}, {Sooley},
  {Sornarajah}, {Streicher}, {Teuben}, {Thomas}, {Tremblay}, {Turner},
  {Terr{\'o}n}, {van Kerkwijk}, {de la Vega}, {Watkins}, {Weaver}, {Whitmore},
  {Woillez}, {Zabalza}, \& {Astropy Contributors}}]{Astropy:2018}
{Astropy Collaboration}, {Price-Whelan}, A.~M., {Sip{\H o}cz}, B.~M., {et~al.}
  2018, \aj, 156, 123

\bibitem[{{Bailer-Jones} {et~al.}(2018){Bailer-Jones}, {Rybizki}, {Fouesneau},
  {Mantelet}, \& {Andrae}}]{BailerJones:2018}
{Bailer-Jones}, C.~A.~L., {Rybizki}, J., {Fouesneau}, M., {Mantelet}, G., \&
  {Andrae}, R. 2018, \aj, 156, 58

\bibitem[{{Beck} {et~al.}(2003){Beck}, {Shukurov}, {Sokoloff}, \&
  {Wielebinski}}]{Beck:2003}
{Beck}, R., {Shukurov}, A., {Sokoloff}, D., \& {Wielebinski}, R. 2003, \aap,
  411, 99

\bibitem[{{Berdyugin} {et~al.}(2014){Berdyugin}, {Piirola}, \&
  {Teerikorpi}}]{Berdyugin:2014}
{Berdyugin}, A., {Piirola}, V., \& {Teerikorpi}, P. 2014, \aap, 561, A24

\bibitem[{{BICEP2/Keck and Planck Collaborations} {et~al.}(2015){BICEP2/Keck
  and Planck Collaborations}, {Ade}, {Aghanim}, {Ahmed}, {Aikin}, {Alexander},
  {Arnaud}, {Aumont}, {Baccigalupi}, \& et~al.}]{BICEPPlanckJoint2015}
{BICEP2/Keck and Planck Collaborations}, {Ade}, P.~A.~R., {Aghanim}, N.,
  {et~al.} 2015, Physical Review Letters, 114, 101301

\bibitem[{{Boulanger} {et~al.}(1996){Boulanger}, {Abergel}, {Bernard},
  {Burton}, {Desert}, {Hartmann}, {Lagache}, \& {Puget}}]{Boulanger:1996}
{Boulanger}, F., {Abergel}, A., {Bernard}, J.-P., {et~al.} 1996, \aap, 312, 256

\bibitem[{{Boulanger} {et~al.}(2018){Boulanger}, {En{\ss}lin}, {Fletcher},
  {Girichides}, {Hackstein}, {Haverkorn}, {H{\"o}randel}, {Jaffe}, {Jasche},
  {Kachelrie{\ss}}, {Kotera}, {Pfrommer}, {Rachen}, {Rodrigues},
  {Ruiz-Granados}, {Seta}, {Shukurov}, {Sigl}, {Steininger}, {Vacca}, {van der
  Velden}, {van Vliet}, \& {Wang}}]{Boulanger:2018}
{Boulanger}, F., {En{\ss}lin}, T., {Fletcher}, A., {et~al.} 2018, Journal of
  Cosmology and Astro-Particle Physics, 2018, 049

\bibitem[{{Burkhart} {et~al.}(2009){Burkhart}, {Falceta-Gon{\c c}alves},
  {Kowal}, \& {Lazarian}}]{Burkhart:2009}
{Burkhart}, B., {Falceta-Gon{\c c}alves}, D., {Kowal}, G., \& {Lazarian}, A.
  2009, \apj, 693, 250

\bibitem[{{Burstein} \& {Heiles}(1978)}]{Burstein:1978}
{Burstein}, D., \& {Heiles}, C. 1978, \apj, 225, 40

\bibitem[{Cabral \& Leedom(1993)}]{Cabral:1993}
Cabral, B., \& Leedom, L.~C. 1993, SIGGRAPH '93 Proceedings of the 20th annual
  conference on Computer graphics and interactive techniques, 263

\bibitem[{{Caldwell} {et~al.}(2017){Caldwell}, {Hirata}, \&
  {Kamionkowski}}]{Caldwell:2017}
{Caldwell}, R.~R., {Hirata}, C., \& {Kamionkowski}, M. 2017, \apj, 839, 91

\bibitem[{{Capitanio} {et~al.}(2017){Capitanio}, {Lallement}, {Vergely},
  {Elyajouri}, \& {Monreal-Ibero}}]{Capitanio:2017}
{Capitanio}, L., {Lallement}, R., {Vergely}, J.~L., {Elyajouri}, M., \&
  {Monreal-Ibero}, A. 2017, \aap, 606, A65

\bibitem[{{Chiang} \& {M{\'e}nard}(2019)}]{Chiang:2019}
{Chiang}, Y.-K., \& {M{\'e}nard}, B. 2019, \apj, 870, 120

\bibitem[{{Clark}(2018)}]{Clark:2018}
{Clark}, S.~E. 2018, \apjl, 857, L10

\bibitem[{{Clark} {et~al.}(2015){Clark}, {Hill}, {Peek}, {Putman}, \&
  {Babler}}]{Clark:2015}
{Clark}, S.~E., {Hill}, J.~C., {Peek}, J.~E.~G., {Putman}, M.~E., \& {Babler},
  B.~L. 2015, Physical Review Letters, 115, 241302

\bibitem[{{Clark} {et~al.}(2019){Clark}, {Peek}, \&
  {Miville-Desch{\^e}nes}}]{Clark:2019}
{Clark}, S.~E., {Peek}, J.~E.~G., \& {Miville-Desch{\^e}nes}, M.-A. 2019, \apj,
  874, 171

\bibitem[{{Clark} {et~al.}(2014){Clark}, {Peek}, \& {Putman}}]{Clark:2014}
{Clark}, S.~E., {Peek}, J.~E.~G., \& {Putman}, M.~E. 2014, \apj, 789, 82

\bibitem[{{Cox}(2005)}]{Cox:2005}
{Cox}, D.~P. 2005, \araa, 43, 337

\bibitem[{{Cox} \& {Reynolds}(1987)}]{Cox:1987}
{Cox}, D.~P., \& {Reynolds}, R.~J. 1987, \araa, 25, 303

\bibitem[{{Davis} \& {Greenstein}(1951)}]{Davis:1951}
{Davis}, Jr., L., \& {Greenstein}, J.~L. 1951, \apj, 114, 206

\bibitem[{{Delabrouille} {et~al.}(2013){Delabrouille}, {Betoule}, {Melin},
  {Miville-Desch{\^e}nes}, {Gonzalez-Nuevo}, {Le Jeune}, {Castex}, {de Zotti},
  {Basak}, {Ashdown}, {Aumont}, {Baccigalupi}, {Banday}, {Bernard}, {Bouchet},
  {Clements}, {da Silva}, {Dickinson}, {Dodu}, {Dolag}, {Elsner}, {Fauvet},
  {Fa{\"y}}, {Giardino}, {Leach}, {Lesgourgues}, {Liguori},
  {Mac{\'{\i}}as-P{\'e}rez}, {Massardi}, {Matarrese}, {Mazzotta}, {Montier},
  {Mottet}, {Paladini}, {Partridge}, {Piffaretti}, {Prezeau}, {Prunet},
  {Ricciardi}, {Roman}, {Schaefer}, \& {Toffolatti}}]{Delabrouille:2013}
{Delabrouille}, J., {Betoule}, M., {Melin}, J.-B., {et~al.} 2013, \aap, 553,
  A96

\bibitem[{{Draine} \& {Fraisse}(2009)}]{Draine:2009}
{Draine}, B.~T., \& {Fraisse}, A.~A. 2009, \apj, 696, 1

\bibitem[{{Dunkley} {et~al.}(2009){Dunkley}, {Amblard}, {Baccigalupi},
  {Betoule}, {Chuss}, {Cooray}, {Delabrouille}, {Dickinson}, {Dobler},
  {Dotson}, {Eriksen}, {Finkbeiner}, {Fixsen}, {Fosalba}, {Fraisse}, {Hirata},
  {Kogut}, {Kristiansen}, {Lawrence}, {Magalha\~{}Es}, {Miville-Deschenes},
  {Meyer}, {Miller}, {Naess}, {Page}, {Peiris}, {Phillips}, {Pierpaoli},
  {Rocha}, {Vaillancourt}, \& {Verde}}]{Dunkley:2009}
{Dunkley}, J., {Amblard}, A., {Baccigalupi}, C., {et~al.} 2009, in American
  Institute of Physics Conference Series, Vol. 1141, American Institute of
  Physics Conference Series, ed. S.~{Dodelson}, D.~{Baumann}, A.~{Cooray},
  J.~{Dunkley}, A.~{Fraisse}, M.~G. {Jackson}, A.~{Kogut}, L.~{Krauss},
  M.~{Zaldarriaga}, \& K.~{Smith}, 222--264

\bibitem[{{Durand} \& {Greenwood}(1958)}]{Durand:1958}
{Durand}, D., \& {Greenwood}, J.~A. 1958, Journal of Geology, 66, 229

\bibitem[{{Falceta-Gon{\c c}alves} {et~al.}(2008){Falceta-Gon{\c c}alves},
  {Lazarian}, \& {Kowal}}]{FalcetaGoncalves+etal_2008}
{Falceta-Gon{\c c}alves}, D., {Lazarian}, A., \& {Kowal}, G. 2008, \apj, 679,
  537

\bibitem[{{Ferri{\`e}re}(2001)}]{Ferriere:2001}
{Ferri{\`e}re}, K.~M. 2001, Reviews of Modern Physics, 73, 1031

\bibitem[{{Fissel} {et~al.}(2016){Fissel}, {Ade}, {Angil{\`e}}, {Ashton},
  {Benton}, {Devlin}, {Dober}, {Fukui}, {Galitzki}, {Gandilo}, {Klein},
  {Korotkov}, {Li}, {Martin}, {Matthews}, {Moncelsi}, {Nakamura},
  {Netterfield}, {Novak}, {Pascale}, {Poidevin}, {Santos}, {Savini}, {Scott},
  {Shariff}, {Diego Soler}, {Thomas}, {Tucker}, {Tucker}, \&
  {Ward-Thompson}}]{Fissel:2016}
{Fissel}, L.~M., {Ade}, P. A.~R., {Angil{\`e}}, F.~E., {et~al.} 2016, \apj,
  824, 134

\bibitem[{{Flauger} {et~al.}(2014){Flauger}, {Hill}, \&
  {Spergel}}]{Flauger:2014}
{Flauger}, R., {Hill}, J.~C., \& {Spergel}, D.~N. 2014, \jcap, 8, 039

\bibitem[{{Gaia Collaboration} {et~al.}(2016){Gaia Collaboration}, {Prusti},
  {de Bruijne}, {Brown}, {Vallenari}, {Babusiaux}, {Bailer-Jones}, {Bastian},
  {Biermann}, {Evans}, \& et~al.}]{Gaia:2016}
{Gaia Collaboration}, {Prusti}, T., {de Bruijne}, J.~H.~J., {et~al.} 2016,
  \aap, 595, A1

\bibitem[{{Gaia Collaboration} {et~al.}(2018){Gaia Collaboration}, {Brown},
  {Vallenari}, {Prusti}, {de Bruijne}, {Babusiaux}, {Bailer-Jones}, {Biermann},
  {Evans}, {Eyer}, \& et~al.}]{GaiaDR2:2018}
{Gaia Collaboration}, {Brown}, A.~G.~A., {Vallenari}, A., {et~al.} 2018, \aap,
  616, A1

\bibitem[{{Ghosh} {et~al.}(2017){Ghosh}, {Boulanger}, {Martin}, {Bracco},
  {Vansyngel}, {Aumont}, {Bock}, {Dor{\'e}}, {Haud}, {Kalberla}, \&
  {Serra}}]{Ghosh:2017}
{Ghosh}, T., {Boulanger}, F., {Martin}, P.~G., {et~al.} 2017, \aap, 601, A71

\bibitem[{{Goldsmith} {et~al.}(2008){Goldsmith}, {Heyer}, {Narayanan}, {Snell},
  {Li}, \& {Brunt}}]{Goldsmith:2008}
{Goldsmith}, P.~F., {Heyer}, M., {Narayanan}, G., {et~al.} 2008, \apj, 680, 428

\bibitem[{{G{\'o}rski} {et~al.}(2005){G{\'o}rski}, {Hivon}, {Banday},
  {Wandelt}, {Hansen}, {Reinecke}, \& {Bartelmann}}]{Healpix:2005}
{G{\'o}rski}, K.~M., {Hivon}, E., {Banday}, A.~J., {et~al.} 2005, \apj, 622,
  759

\bibitem[{{Grain} {et~al.}(2009){Grain}, {Tristram}, \& {Stompor}}]{Grain:2009}
{Grain}, J., {Tristram}, M., \& {Stompor}, R. 2009, \prd, 79, 123515

\bibitem[{{Green} {et~al.}(2019){Green}, {Schlafly}, {Zucker}, {Speagle}, \&
  {Finkbeiner}}]{Green:2019}
{Green}, G.~M., {Schlafly}, E.~F., {Zucker}, C., {Speagle}, J.~S., \&
  {Finkbeiner}, D.~P. 2019, arXiv e-prints, arXiv:1905.02734

\bibitem[{{Guillet} {et~al.}(2018){Guillet}, {Fanciullo}, {Verstraete},
  {Boulanger}, {Jones}, {Miville-Desch{\^e}nes}, {Ysard}, {Levrier}, \&
  {Alves}}]{Guillet:2018}
{Guillet}, V., {Fanciullo}, L., {Verstraete}, L., {et~al.} 2018, \aap, 610, A16

\bibitem[{{Haverkorn}(2015)}]{Haverkorn:2015}
{Haverkorn}, M. 2015, in Astrophysics and Space Science Library, Vol. 407,
  Magnetic Fields in Diffuse Media, ed. A.~{Lazarian}, E.~M. {de Gouveia Dal
  Pino}, \& C.~{Melioli}, 483

\bibitem[{{Haverkorn} {et~al.}(2019){Haverkorn}, {Boulanger}, {En{\ss}lin},
  {H{\"o}randel}, {Jaffe}, {Jasche}, {Rachen}, \& {Shukurov}}]{Haverkorn:2019}
{Haverkorn}, M., {Boulanger}, F., {En{\ss}lin}, T., {et~al.} 2019, Galaxies, 7,
  17

\bibitem[{{Heiles}(1984)}]{Heiles:1984}
{Heiles}, C. 1984, \apjs, 55, 585

\bibitem[{{Heiles}(2000)}]{Heiles:2000}
---. 2000, \aj, 119, 923

\bibitem[{{Hensley} \& {Bull}(2018)}]{HensleyBull:2018}
{Hensley}, B.~S., \& {Bull}, P. 2018, \apj, 853, 127

\bibitem[{{Hensley} {et~al.}(2019){Hensley}, {Zhang}, \& {Bock}}]{Hensley:2019}
{Hensley}, B.~S., {Zhang}, C., \& {Bock}, J.~J. 2019, arXiv e-prints,
  arXiv:1909.07394

\bibitem[{{HI4PI Collaboration} {et~al.}(2016){HI4PI Collaboration}, {Ben
  Bekhti}, {Fl{\"o}er}, {Keller}, {Kerp}, {Lenz}, {Winkel}, {Bailin},
  {Calabretta}, {Dedes}, {Ford}, {Gibson}, {Haud}, {Janowiecki}, {Kalberla},
  {Lockman}, {McClure-Griffiths}, {Murphy}, {Nakanishi}, {Pisano}, \&
  {Staveley-Smith}}]{HI4PI:2016}
{HI4PI Collaboration}, {Ben Bekhti}, N., {Fl{\"o}er}, L., {et~al.} 2016, \aap,
  594, A116

\bibitem[{Hough(1962)}]{Hough:1962tb}
Hough, P. V.~C. 1962, {Method and Means for Recognizing Complex Patterns}, US
  Patent

\bibitem[{{Huffenberger} {et~al.}(2019){Huffenberger}, {Rotti}, \&
  {Collins}}]{Huffenberger:2019}
{Huffenberger}, K.~M., {Rotti}, A., \& {Collins}, D.~C. 2019, arXiv e-prints,
  arXiv:1906.10052

\bibitem[{{Hunter}(2007)}]{Matplotlib:2007}
{Hunter}, J.~D. 2007, Computing in Science and Engineering, 9, 90

\bibitem[{{Hutschenreuter} \& {En{\ss}lin}(2019)}]{Hutschenreuter:2019}
{Hutschenreuter}, S., \& {En{\ss}lin}, T.~A. 2019, arXiv e-prints,
  arXiv:1903.06735

\bibitem[{Inoue \& Inutsuka(2016)}]{Inoue:2016wb}
Inoue, T., \& Inutsuka, S.-i. 2016, \apj, 833, 10

\bibitem[{{Jaffe}(2019)}]{Jaffe:2019}
{Jaffe}, T.~R. 2019, Galaxies, 7, 52

\bibitem[{{Jaffe} {et~al.}(2013){Jaffe}, {Ferri{\`e}re}, {Banday}, {Strong},
  {Orlando}, {Mac{\'\i}as-P{\'e}rez}, {Fauvet}, {Combet}, \&
  {Falgarone}}]{Jaffe:2013}
{Jaffe}, T.~R., {Ferri{\`e}re}, K.~M., {Banday}, A.~J., {et~al.} 2013, \mnras,
  431, 683

\bibitem[{{Jansson} \& {Farrar}(2012)}]{JanssonFarrar:2012}
{Jansson}, R., \& {Farrar}, G.~R. 2012, \apj, 757, 14

\bibitem[{{Jeli{\'c}} {et~al.}(2018){Jeli{\'c}}, {Prelogovi{\'c}}, {Haverkorn},
  {Remeijn}, \& {Klind{\v z}i{\'c}}}]{Jelic:2018}
{Jeli{\'c}}, V., {Prelogovi{\'c}}, D., {Haverkorn}, M., {Remeijn}, J., \&
  {Klind{\v z}i{\'c}}, D. 2018, \aap, 615, L3

\bibitem[{{Jeli{\'c}} {et~al.}(2015){Jeli{\'c}}, {de Bruyn}, {Pandey},
  {Mevius}, {Haverkorn}, {Brentjens}, {Koopmans}, {Zaroubi}, {Abdalla}, {Asad},
  {Bus}, {Chapman}, {Ciardi}, {Fernandez}, {Ghosh}, {Harker}, {Iliev},
  {Jensen}, {Kazemi}, {Mellema}, {Offringa}, {Patil}, {Vedantham}, \&
  {Yatawatta}}]{Jelic:2015}
{Jeli{\'c}}, V., {de Bruyn}, A.~G., {Pandey}, V.~N., {et~al.} 2015, \aap, 583,
  A137

\bibitem[{{Jow} {et~al.}(2018){Jow}, {Hill}, {Scott}, {Soler}, {Martin},
  {Devlin}, {Fissel}, \& {Poidevin}}]{Jow:2018}
{Jow}, D.~L., {Hill}, R., {Scott}, D., {et~al.} 2018, \mnras, 474, 1018

\bibitem[{{Kalberla} {et~al.}(2005){Kalberla}, {Burton}, {Hartmann}, {Arnal},
  {Bajaja}, {Morras}, \& {P{\"o}ppel}}]{Kalberla:2005}
{Kalberla}, P.~M.~W., {Burton}, W.~B., {Hartmann}, D., {et~al.} 2005, \aap,
  440, 775

\bibitem[{{Kalberla} \& {Kerp}(2016)}]{KalberlaKerp:2016}
{Kalberla}, P.~M.~W., \& {Kerp}, J. 2016, \aap, 595, A37

\bibitem[{{Kalberla} {et~al.}(2016){Kalberla}, {Kerp}, {Haud}, {Winkel}, {Ben
  Bekhti}, {Fl{\"o}er}, \& {Lenz}}]{Kalberla:2016}
{Kalberla}, P.~M.~W., {Kerp}, J., {Haud}, U., {et~al.} 2016, \apj, 821, 117

\bibitem[{{Kamionkowski} {et~al.}(1997){Kamionkowski}, {Kosowsky}, \&
  {Stebbins}}]{Kamionkowski:1997}
{Kamionkowski}, M., {Kosowsky}, A., \& {Stebbins}, A. 1997, \prl, 78, 2058

\bibitem[{{Kim} {et~al.}(2019){Kim}, {Choi}, \& {Flauger}}]{Kim:2019}
{Kim}, C.-G., {Choi}, S.~K., \& {Flauger}, R. 2019, \apj, 880, 106

\bibitem[{{King} {et~al.}(2018){King}, {Fissel}, {Chen}, \& {Li}}]{King:2018}
{King}, P.~K., {Fissel}, L.~M., {Chen}, C.-Y., \& {Li}, Z.-Y. 2018, \mnras,
  474, 5122

\bibitem[{{Kritsuk} {et~al.}(2018){Kritsuk}, {Flauger}, \&
  {Ustyugov}}]{Kritsuk:2018}
{Kritsuk}, A.~G., {Flauger}, R., \& {Ustyugov}, S.~D. 2018, Physical Review
  Letters, 121, 021104

\bibitem[{{Lallement} {et~al.}(2014){Lallement}, {Vergely}, {Valette},
  {Puspitarini}, {Eyer}, \& {Casagrande}}]{Lallement:2014}
{Lallement}, R., {Vergely}, J.-L., {Valette}, B., {et~al.} 2014, \aap, 561, A91

\bibitem[{{Lallement} {et~al.}(2003){Lallement}, {Welsh}, {Vergely}, {Crifo},
  \& {Sfeir}}]{Lallement:2003}
{Lallement}, R., {Welsh}, B.~Y., {Vergely}, J.~L., {Crifo}, F., \& {Sfeir}, D.
  2003, \aap, 411, 447

\bibitem[{{Lee} \& {Draine}(1985)}]{Lee:1985}
{Lee}, H.~M., \& {Draine}, B.~T. 1985, \apj, 290, 211

\bibitem[{{Lenz} {et~al.}(2019){Lenz}, {Dor{\'e}}, \& {Lagache}}]{Lenz:2019}
{Lenz}, D., {Dor{\'e}}, O., \& {Lagache}, G. 2019, arXiv e-prints,
  arXiv:1905.00426

\bibitem[{{Lenz} {et~al.}(2017){Lenz}, {Hensley}, \& {Dor{\'e}}}]{Lenz:2017}
{Lenz}, D., {Hensley}, B.~S., \& {Dor{\'e}}, O. 2017, \apj, 846, 38

\bibitem[{{Lloyd} \& {Harwit}(1973)}]{Lloyd:1973}
{Lloyd}, S., \& {Harwit}, M.~O. 1973, in IAU Symposium, Vol.~52, Interstellar
  Dust and Related Topics, ed. J.~M. {Greenberg} \& H.~C. {van de Hulst}, 203

\bibitem[{Malinen {et~al.}(2016)Malinen, Montier, Montillaud, Juvela,
  Ristorcelli, Clark, Bern{\'e}, Bernard, Pelkonen, \& Collins}]{Malinen:2016}
Malinen, J., Montier, L., Montillaud, J., {et~al.} 2016, \mnras, 460, 1934

\bibitem[{{Marchal} {et~al.}(2019){Marchal}, {Miville-Desch{\^e}nes}, {Orieux},
  {Gac}, {Soussen}, {Lesot}, {d'Allonnes}, \& {Salom{\'e}}}]{Marchal:2019}
{Marchal}, A., {Miville-Desch{\^e}nes}, M.-A., {Orieux}, F., {et~al.} 2019,
  \aap, 626, A101

\bibitem[{{Marelli} {et~al.}(2019){Marelli}, {Tiengo}, {De Luca}, {Mignani},
  {Salvetti}, {Saz Parkinson}, \& {Lisini}}]{Marelli:2019}
{Marelli}, M., {Tiengo}, A., {De Luca}, A., {et~al.} 2019, Astronomy and
  Astrophysics, 624, A53

\bibitem[{{Marshall} {et~al.}(2006){Marshall}, {Robin}, {Reyl{\'e}},
  {Schultheis}, \& {Picaud}}]{Marshall:2006}
{Marshall}, D.~J., {Robin}, A.~C., {Reyl{\'e}}, C., {Schultheis}, M., \&
  {Picaud}, S. 2006, \aap, 453, 635

\bibitem[{{Martin} {et~al.}(2015){Martin}, {Blagrave}, {Lockman}, {Pinheiro
  Gon{\c c}alves}, {Boothroyd}, {Joncas}, {Miville-Desch{\^e}nes}, \&
  {Stephan}}]{Martin:2015}
{Martin}, P.~G., {Blagrave}, K.~P.~M., {Lockman}, F.~J., {et~al.} 2015, \apj,
  809, 153

\bibitem[{{Mart{\'{\i}}nez-Solaeche} {et~al.}(2018){Mart{\'{\i}}nez-Solaeche},
  {Karakci}, \& {Delabrouille}}]{MartinezSolaeche:2018}
{Mart{\'{\i}}nez-Solaeche}, G., {Karakci}, A., \& {Delabrouille}, J. 2018,
  \mnras, 476, 1310

\bibitem[{{McClure-Griffiths} {et~al.}(2006){McClure-Griffiths}, {Dickey},
  {Gaensler}, {Green}, \& {Haverkorn}}]{McClure-Griffiths:2006}
{McClure-Griffiths}, N.~M., {Dickey}, J.~M., {Gaensler}, B.~M., {Green}, A.~J.,
  \& {Haverkorn}, M. 2006, \apj, 652, 1339

\bibitem[{{McClure-Griffiths} {et~al.}(2009){McClure-Griffiths}, {Pisano},
  {Calabretta}, {Ford}, {Lockman}, {Staveley-Smith}, {Kalberla}, {Bailin},
  {Dedes}, {Janowiecki}, {Gibson}, {Murphy}, {Nakanishi}, \&
  {Newton-McGee}}]{McClure-Griffiths:2009}
{McClure-Griffiths}, N.~M., {Pisano}, D.~J., {Calabretta}, M.~R., {et~al.}
  2009, \apjs, 181, 398

\bibitem[{{Murray} {et~al.}(2018){Murray}, {Peek}, {Lee}, \&
  {Stanimirovi{\'c}}}]{Murray:2018}
{Murray}, C.~E., {Peek}, J.~E.~G., {Lee}, M.-Y., \& {Stanimirovi{\'c}}, S.
  2018, \apj, 862, 131

\bibitem[{Oliphant(2015)}]{Oliphant:2015:GN:2886196}
Oliphant, T.~E. 2015, Guide to NumPy, 2nd edn. (USA: CreateSpace Independent
  Publishing Platform)

\bibitem[{{Panopoulou} {et~al.}(2016){Panopoulou}, {Tassis}, {Blinov},
  {Pavlidou}, {King}, {Paleologou}, {Ramaprakash}, {Angelakis},
  {Balokovi{\'c}}, {Das}, {Feiler}, {Hovatta}, {Khodade}, {Kiehlmann}, {Kus},
  {Kylafis}, {Liodakis}, {Mahabal}, {Modi}, {Myserlis}, {Papadakis},
  {Papamastorakis}, {Pazderska}, {Pazderski}, {Pearson}, {Rajarshi},
  {Readhead}, {Reig}, \& {Zensus}}]{Panopoulou2016Erratum}
{Panopoulou}, G., {Tassis}, K., {Blinov}, D., {et~al.} 2016, \mnras, 462, 2011

\bibitem[{{Panopoulou} {et~al.}(2019){Panopoulou}, {Tassis}, {Skalidis},
  {Blinov}, {Liodakis}, {Pavlidou}, {Potter}, {Ramaprakash}, {Readhead}, \&
  {Wehus}}]{Panopoulou:2019Tomo}
{Panopoulou}, G.~V., {Tassis}, K., {Skalidis}, R., {et~al.} 2019, \apj, 872, 56

\bibitem[{{Peek} \& {Clark}(2019)}]{PeekClark:2019}
{Peek}, J.~E.~G., \& {Clark}, S.~E. 2019, arXiv e-prints, arXiv:1909.09647

\bibitem[{{Peek} {et~al.}(2018){Peek}, {Babler}, {Zheng}, {Clark}, {Douglas},
  {Korpela}, {Putman}, {Stanimirovi{\'c}}, {Gibson}, \& {Heiles}}]{Peek:2018}
{Peek}, J.~E.~G., {Babler}, B.~L., {Zheng}, Y., {et~al.} 2018, \apjs, 234, 2

\bibitem[{{Planck Collaboration Int. XIX}(2015)}]{PlanckXIX}
{Planck Collaboration Int. XIX}. 2015, \aap, 576, A104

\bibitem[{{Planck Collaboration Int. XLIV}(2016)}]{PlanckXLIV}
{Planck Collaboration Int. XLIV}. 2016, \aap, 596, A105

\bibitem[{{Planck Collaboration Int. XX}(2015)}]{Planck_Int_XX}
{Planck Collaboration Int. XX}. 2015, \aap, 576, A105

\bibitem[{{Planck Collaboration Int. XXX}(2016)}]{Planck_Int_XXX}
{Planck Collaboration Int. XXX}. 2016, \aap, 586, A133

\bibitem[{{Planck Collaboration Int. XXXII}(2016)}]{PlanckXXXII}
{Planck Collaboration Int. XXXII}. 2016, \aap, 586, A135

\bibitem[{{Planck Collaboration Int. XXXVIII}(2016)}]{PlanckXXXVIII}
{Planck Collaboration Int. XXXVIII}. 2016, \aap, 586, A141

\bibitem[{{Planck Collaboration XI}(2018)}]{Planck2018XI}
{Planck Collaboration XI}. 2018, arXiv e-prints, arXiv:1801.04945

\bibitem[{{Planck Collaboration XII}(2018)}]{Planck2018XII}
{Planck Collaboration XII}. 2018, ArXiv e-prints, arXiv:1807.06212

\bibitem[{{Planck Collaboration XXIV}(2011)}]{Planck2011XXIV}
{Planck Collaboration XXIV}. 2011, \aap, 536, A24

\bibitem[{{Plaszczynski} {et~al.}(2014){Plaszczynski}, {Montier}, {Levrier}, \&
  {Tristram}}]{Plaszczynski:2014}
{Plaszczynski}, S., {Montier}, L., {Levrier}, F., \& {Tristram}, M. 2014,
  \mnras, 439, 4048

\bibitem[{{Poh} \& {Dodelson}(2017)}]{Poh:2017}
{Poh}, J., \& {Dodelson}, S. 2017, \prd, 95, 103511

\bibitem[{{Putman} {et~al.}(2012){Putman}, {Peek}, \& {Joung}}]{Putman:2012}
{Putman}, M.~E., {Peek}, J.~E.~G., \& {Joung}, M.~R. 2012, \araa, 50, 491

\bibitem[{{Ramaprakash} {et~al.}(2019){Ramaprakash}, {Rajarshi}, {Das},
  {Khodade}, {Modi}, {Panopoulou}, {Maharana}, {Blinov}, {Angelakis},
  {Casadio}, {Fuhrmann}, {Hovatta}, {Kiehlmann}, {King}, {Kylafis},
  {Kougentakis}, {Kus}, {Mahabal}, {Marecki}, {Myserlis}, {Paterakis},
  {Paleologou}, {Liodakis}, {Papadakis}, {Papamastorakis}, {Pavlidou},
  {Pazderski}, {Pearson}, {Readhead}, {Reig}, {S{\l}owikowska}, {Tassis}, \&
  {Zensus}}]{Ramaprakash:2019}
{Ramaprakash}, A.~N., {Rajarshi}, C.~V., {Das}, H.~K., {et~al.} 2019, \mnras,
  485, 2355

\bibitem[{{Remazeilles} {et~al.}(2011){Remazeilles}, {Delabrouille}, \&
  {Cardoso}}]{Remazeilles:2011}
{Remazeilles}, M., {Delabrouille}, J., \& {Cardoso}, J.-F. 2011, \mnras, 418,
  467

\bibitem[{{Riener} {et~al.}(2019){Riener}, {Kainulainen}, {Henshaw}, {Orkisz},
  {Murray}, \& {Beuther}}]{Riener:2019}
{Riener}, M., {Kainulainen}, J., {Henshaw}, J.~D., {et~al.} 2019, \aap, 628,
  A78

\bibitem[{{Robishaw} \& {Heiles}(2018)}]{Robishaw:2018}
{Robishaw}, T., \& {Heiles}, C. 2018, arXiv e-prints, arXiv:1806.07391

\bibitem[{{Rotti} \& {Huffenberger}(2019)}]{Rotti:2019}
{Rotti}, A., \& {Huffenberger}, K. 2019, \jcap, 1, 045

\bibitem[{{Schad}(2017)}]{Schad:2017}
{Schad}, T. 2017, \solphys, 292, 132

\bibitem[{Seljak \& Zaldarriaga(1997)}]{Seljak:1997}
Seljak, U. b.~u., \& Zaldarriaga, M. 1997, Phys. Rev. Lett., 78, 2054

\bibitem[{{Smith}(2006)}]{Smith:2006}
{Smith}, K.~M. 2006, \nar, 50, 1025

\bibitem[{Soler {et~al.}(2013)Soler, Hennebelle, Martin, Miville-Desch{\^e}nes,
  Netterfield, \& Fissel}]{Soler:2013}
Soler, J.~D., Hennebelle, P., Martin, P.~G., {et~al.} 2013, \apj, 774, 128

\bibitem[{{Tassis} \& {Pavlidou}(2015)}]{TassisPavlidou:2015}
{Tassis}, K., \& {Pavlidou}, V. 2015, \mnras, 451, L90

\bibitem[{{Tassis} {et~al.}(2018){Tassis}, {Ramaprakash}, {Readhead}, {Potter},
  {Wehus}, {Panopoulou}, {Blinov}, {Eriksen}, {Hensley}, {Karakci},
  {Kypriotakis}, {Maharana}, {Ntormousi}, {Pavlidou}, {Pearson}, \&
  {Skalidis}}]{PASIPHAE:2018}
{Tassis}, K., {Ramaprakash}, A.~N., {Readhead}, A.~C.~S., {et~al.} 2018, arXiv
  e-prints, arXiv:1810.05652

\bibitem[{{Tchernyshyov} \& {Peek}(2017)}]{Tchernyshyov:2017}
{Tchernyshyov}, K., \& {Peek}, J.~E.~G. 2017, \aj, 153, 8

\bibitem[{{Terral} \& {Ferri{\`e}re}(2017)}]{Terral:2017}
{Terral}, P., \& {Ferri{\`e}re}, K. 2017, \aap, 600, A29

\bibitem[{{Thomson} {et~al.}(2019){Thomson}, {Landecker}, {Dickey},
  {McClure-Griffiths}, {Wolleben}, {Carretti}, {Fletcher}, {Federrath}, {Hill},
  {Mao}, {Gaensler}, {Haverkorn}, {Clark}, {Van Eck}, \& {West}}]{Thomson:2019}
{Thomson}, A.~J.~M., {Landecker}, T.~L., {Dickey}, J.~M., {et~al.} 2019,
  \mnras, arXiv:1905.09285

\bibitem[{{Thorne} {et~al.}(2017){Thorne}, {Dunkley}, {Alonso}, \&
  {N{\ae}ss}}]{Thorne:2017}
{Thorne}, B., {Dunkley}, J., {Alonso}, D., \& {N{\ae}ss}, S. 2017, \mnras, 469,
  2821

\bibitem[{{Thorne} {et~al.}(2019){Thorne}, {Dunkley}, {Alonso}, {Abitbol},
  {Errard}, {Hill}, {Keating}, {Teply}, \& {Wollack}}]{Thorne:2019}
{Thorne}, B., {Dunkley}, J., {Alonso}, D., {et~al.} 2019, arXiv e-prints,
  arXiv:1905.08888

\bibitem[{{Tingay} {et~al.}(2013){Tingay}, {Goeke}, {Bowman}, {Emrich}, {Ord},
  {Mitchell}, {Morales}, {Booler}, {Crosse}, {Wayth}, {Lonsdale}, {Tremblay},
  {Pallot}, {Colegate}, {Wicenec}, {Kudryavtseva}, {Arcus}, {Barnes},
  {Bernardi}, {Briggs}, {Burns}, {Bunton}, {Cappallo}, {Corey}, {Deshpande},
  {Desouza}, {Gaensler}, {Greenhill}, {Hall}, {Hazelton}, {Herne}, {Hewitt},
  {Johnston-Hollitt}, {Kaplan}, {Kasper}, {Kincaid}, {Koenig}, {Kratzenberg},
  {Lynch}, {Mckinley}, {Mcwhirter}, {Morgan}, {Oberoi}, {Pathikulangara},
  {Prabu}, {Remillard}, {Rogers}, {Roshi}, {Salah}, {Sault}, {Udaya-Shankar},
  {Schlagenhaufer}, {Srivani}, {Stevens}, {Subrahmanyan}, {Waterson},
  {Webster}, {Whitney}, {Williams}, {Williams}, \& {Wyithe}}]{Tingay:2013}
{Tingay}, S.~J., {Goeke}, R., {Bowman}, J.~D., {et~al.} 2013, \pasa, 30, e007

\bibitem[{{Van Eck} {et~al.}(2017){Van Eck}, {Haverkorn}, {Alves}, {Beck}, {de
  Bruyn}, {En{\ss}lin}, {Farnes}, {Ferri{\`e}re}, {Heald}, {Horellou},
  {Horneffer}, {Iacobelli}, {Jeli{\'c}}, {Mart{\'{\i}}-Vidal}, {Mulcahy},
  {Reich}, {R{\"o}ttgering}, {Scaife}, {Schnitzeler}, {Sobey}, \&
  {Sridhar}}]{vanEck:2017}
{Van Eck}, C.~L., {Haverkorn}, M., {Alves}, M.~I.~R., {et~al.} 2017, \aap, 597,
  A98

\bibitem[{{van Haarlem} {et~al.}(2013){van Haarlem}, {Wise}, {Gunst}, {Heald},
  {McKean}, {Hessels}, {de Bruyn}, {Nijboer}, {Swinbank}, {Fallows},
  {Brentjens}, {Nelles}, {Beck}, {Falcke}, {Fender}, {H{\"o}randel},
  {Koopmans}, {Mann}, {Miley}, {R{\"o}ttgering}, {Stappers}, {Wijers},
  {Zaroubi}, {van den Akker}, {Alexov}, {Anderson}, {Anderson}, {van Ardenne},
  {Arts}, {Asgekar}, {Avruch}, {Batejat}, {B{\"a}hren}, {Bell}, {Bell}, {van
  Bemmel}, {Bennema}, {Bentum}, {Bernardi}, {Best}, {B{\^i}rzan}, {Bonafede},
  {Boonstra}, {Braun}, {Bregman}, {Breitling}, {van de Brink}, {Broderick},
  {Broekema}, {Brouw}, {Br{\"u}ggen}, {Butcher}, {van Cappellen}, {Ciardi},
  {Coenen}, {Conway}, {Coolen}, {Corstanje}, {Damstra}, {Davies}, {Deller},
  {Dettmar}, {van Diepen}, {Dijkstra}, {Donker}, {Doorduin}, {Dromer}, {Drost},
  {van Duin}, {Eisl{\"o}ffel}, {van Enst}, {Ferrari}, {Frieswijk}, {Gankema},
  {Garrett}, {de Gasperin}, {Gerbers}, {de Geus}, {Grie{\ss}meier}, {Grit},
  {Gruppen}, {Hamaker}, {Hassall}, {Hoeft}, {Holties}, {Horneffer}, {van der
  Horst}, {van Houwelingen}, {Huijgen}, {Iacobelli}, {Intema}, {Jackson},
  {Jelic}, {de Jong}, {Juette}, {Kant}, {Karastergiou}, {Koers}, {Kollen},
  {Kondratiev}, {Kooistra}, {Koopman}, {Koster}, {Kuniyoshi}, {Kramer},
  {Kuper}, {Lambropoulos}, {Law}, {van Leeuwen}, {Lemaitre}, {Loose}, {Maat},
  {Macario}, {Markoff}, {Masters}, {McFadden}, {McKay-Bukowski}, {Meijering},
  {Meulman}, {Mevius}, {Middelberg}, {Millenaar}, {Miller-Jones}, {Mohan},
  {Mol}, {Morawietz}, {Morganti}, {Mulcahy}, {Mulder}, {Munk}, {Nieuwenhuis},
  {van Nieuwpoort}, {Noordam}, {Norden}, {Noutsos}, {Offringa}, {Olofsson},
  {Omar}, {Orr{\'u}}, {Overeem}, {Paas}, {Pandey-Pommier}, {Pandey}, {Pizzo},
  {Polatidis}, {Rafferty}, {Rawlings}, {Reich}, {de Reijer}, {Reitsma},
  {Renting}, {Riemers}, {Rol}, {Romein}, {Roosjen}, {Ruiter}, {Scaife}, {van
  der Schaaf}, {Scheers}, {Schellart}, {Schoenmakers}, {Schoonderbeek},
  {Serylak}, {Shulevski}, {Sluman}, {Smirnov}, {Sobey}, {Spreeuw}, {Steinmetz},
  {Sterks}, {Stiepel}, {Stuurwold}, {Tagger}, {Tang}, {Tasse}, {Thomas},
  {Thoudam}, {Toribio}, {van der Tol}, {Usov}, {van Veelen}, {van der Veen},
  {ter Veen}, {Verbiest}, {Vermeulen}, {Vermaas}, {Vocks}, {Vogt}, {de Vos},
  {van der Wal}, {van Weeren}, {Weggemans}, {Weltevrede}, {White}, {Wijnholds},
  {Wilhelmsson}, {Wucknitz}, {Yatawatta}, {Zarka}, {Zensus}, \& {van
  Zwieten}}]{vanHaarlem:2013}
{van Haarlem}, M.~P., {Wise}, M.~W., {Gunst}, A.~W., {et~al.} 2013, \aap, 556,
  A2

\bibitem[{{Vansyngel} {et~al.}(2017){Vansyngel}, {Boulanger}, {Ghosh},
  {Wandelt}, {Aumont}, {Bracco}, {Levrier}, {Martin}, \&
  {Montier}}]{Vansyngel:2017}
{Vansyngel}, F., {Boulanger}, F., {Ghosh}, T., {et~al.} 2017, \aap, 603, A62

\bibitem[{{Wakker} \& {Boulanger}(1986)}]{Wakker:1986}
{Wakker}, B.~P., \& {Boulanger}, F. 1986, \aap, 170, 84

\bibitem[{{Winkel} {et~al.}(2016){Winkel}, {Kerp}, {Fl{\"o}er}, {Kalberla},
  {Ben Bekhti}, {Keller}, \& {Lenz}}]{Winkel:2016}
{Winkel}, B., {Kerp}, J., {Fl{\"o}er}, L., {et~al.} 2016, \aap, 585, A41

\bibitem[{{Wolleben} {et~al.}(2019){Wolleben}, {Landecker}, {Carretti},
  {Dickey}, {Fletcher}, {McClure-Griffiths}, {McConnell}, {Thomson}, {Hill},
  {Gaensler}, {Han}, {Haverkorn}, {Leahy}, {Reich}, \&
  {Taylor}}]{Wolleben:2019}
{Wolleben}, M., {Landecker}, T.~L., {Carretti}, E., {et~al.} 2019, \aj, 158, 44

\bibitem[{{Zaroubi} {et~al.}(2015){Zaroubi}, {Jeli{\'c}}, {de Bruyn},
  {Boulanger}, {Bracco}, {Kooistra}, {Alves}, {Brentjens}, {Ferri{\`e}re},
  {Ghosh}, {Koopmans}, {Levrier}, {Miville-Desch{\^e}nes}, {Montier}, {Pandey},
  \& {Soler}}]{Zaroubi:2015}
{Zaroubi}, S., {Jeli{\'c}}, V., {de Bruyn}, A.~G., {et~al.} 2015, \mnras, 454,
  L46

\bibitem[{Zonca {et~al.}(2019)Zonca, Singer, Lenz, Reinecke, Rosset, Hivon, \&
  Gorski}]{Zonca:2019}
Zonca, A., Singer, L., Lenz, D., {et~al.} 2019, Journal of Open Source
  Software, 4, 1298

\end{thebibliography}

\end{document}